%% file: NF_Wideband_PLS.tex
\documentclass[journal]{IEEEtran}
\usepackage{amsmath,amssymb}
\usepackage{subcaption}
\usepackage{graphicx,graphics,color,psfrag}
\usepackage{cite,balance}
\usepackage{caption}
\allowdisplaybreaks
\usepackage{algorithm}
\usepackage{algorithmic}
\usepackage{accents}
\usepackage{amsthm}
\usepackage{bm}
\usepackage{url}
\usepackage[english]{babel}
\usepackage{multirow}
\usepackage{enumerate}
\usepackage{cases}
\usepackage{stfloats}
\usepackage{dsfont}
\usepackage{color,soul}
\usepackage{amsfonts}
\usepackage{cite,graphicx,amsmath,amssymb}
\usepackage{fancyhdr}
\usepackage{hhline}
\usepackage{graphicx,graphics}
\usepackage{array,color}
\usepackage{mathtools}
\usepackage{amsmath}
\usepackage{hyperref}
\hypersetup{hypertex=true,
colorlinks=true,
linkcolor=black,
anchorcolor=black,
citecolor=black}

\include{header}

\begin{document}
\setlength{\textheight}{9.83in}
\setlength{\textwidth}{7.4in}
\captionsetup[figure]{name={Fig.}}
\title{\huge 
Wideband Physical Layer Security in Mixed Near-Field and Far-Field Communications}
\author{Yunpu~Zhang,
Changsheng~You,~\IEEEmembership{Member,~IEEE}, and Hing Cheung So,~\IEEEmembership{Fellow,~IEEE}
\thanks{
Y. Zhang and H. C. So are with the Department of Electrical Engineering, City University of Hong Kong, Hong Kong (e-mail:
yunpu.zhang@my.cityu.edu.hk, hcso@ee.cityu.edu.hk).
C. You is with the Department of Electronic and Electrical Engineering, Southern University of Science and Technology (SUSTech), Shenzhen
518055, China (e-mail: youcs@sustech.edu.cn).
 \emph{(Corresponding
 author: Changsheng~You.)}
}\vspace{-20pt}}
\maketitle

% We consider a mixed-field wideband communication system. 

% \begin{itemize}
%     \item narrowband PLS 
%     \item wideband PLS
%    % \item mixed-field narrowband PLS (?)
%     \item mixed-field wideband PLS
% \end{itemize}

% The difference between the conventional far-field wideband PLS and the mixed-field PLS lies in
% \begin{itemize}
%     \item First, 
% \end{itemize}
\begin{abstract}
 Prior studies on wideband physical layer security (PLS) have mostly focused on either far-field or near-field communication systems. In this paper, we investigate PLS in a more general and practical \emph{mixed near-field and far-field} wideband communication scenario, where a base station (BS) equipped with an extremely large-scale array (XL-array) serves multiple legitimate users in the far field, while a near-field eavesdropper attempts to intercept confidential information at close range. Specifically, we formulate an optimization problem to maximize the secrecy rate of all legitimate users across all subcarriers by jointly designing the transmit beamforming and artificial noise (AN), subject to per-subcarrier power constraints. To gain useful insights into the unique characteristics of mixed-field wideband PLS, we first analyze a special case with a single legitimate user and an eavesdropper. Interestingly, we show that the secure transmission regions in mixed-field wideband systems are not the same across all subcarriers, as revealed by subcarrier-specific \emph{secrecy conditions} derived from the Fresnel integrals. Moreover, we theoretically demonstrate that introducing \emph{frequency-selective} AN into the system offers two key benefits: 1) it transforms originally insecure subcarriers into secure ones, and 2) it substantially improves the overall secrecy performance. For the general multiuser setup, we propose an efficient \emph{two-stage} hybrid beamforming algorithm to jointly design the transmit beamforming and AN, ensuring secure transmission under mixed-field conditions. Building on our analytical insights, we further develop an efficient and low-complexity algorithm that designs the AN beamformers by directly aligning them with the eavesdropper's channel, thereby eliminating the need to optimize AN beamformers across all subcarriers. Finally, numerical results confirm that the secrecy rates across subcarriers decrease as the subcarrier frequency increases. The superior performance of both proposed algorithms over existing benchmark schemes is also demonstrated.
\end{abstract}
\vspace{-14pt}
\section{Introduction}
Extremely large-scale arrays (XL-arrays) have been envisioned as a promising technology to significantly improve the spectral efficiency and spatial resolution of future
wireless systems \cite{10496996,liu2023near,9903389,you2024next,li2025intelligent,lin2024single}.
Notably, the large aperture of XL-arrays fundamentally alters the electromagnetic propagation characteristics, transitioning from conventional far-field communications with planar wavefronts to near-field communications characterized by {spherical wavefronts}, and ultimately to the more general and practical scenario, namely, \emph{mixed near-field and far-field communications}, where both planar and spherical wavefronts coexist. In this context, unlike traditional far-field beam steering, which focuses energy in specific angular directions, spherical wavefronts enable the unique capability of near-field \emph{beam-focusing}, allowing the XL-array to concentrate energy at a specific region. 
This fundamental difference between far-field and near-field propagation gives rise to a distinct \emph{energy-spread} effect: when a far-field beam is steered toward a specific angle, its energy may disperse across multiple directions in the near field. This phenomenon leads to severe mixed-field inter-user interference from the perspective of information transmission \cite{zhang2023mixed}, and mixed-field information leakage from the perspective of secure communications \cite{liu2025physical}. In this paper, we study physical layer security (PLS) in a general and practical \emph{mixed near-field and far-field wideband} communication system. In particular, we show that secure transmission in such mixed-field wideband systems exhibits inherent variations across subcarriers. Furthermore, the introduction of \emph{frequency-selective} artificial noise (AN) is demonstrated to significantly enhance the secrecy performance.
\vspace{-12pt}
\subsection{Prior Work}
% the authors of \cite{10879532} investigated an orthogonal frequency division multiplexing
% (OFDM)-based wideband THz communication system assisted by a simultaneously transmitting and reflecting reconfigurable intelligent surface (STAR-RIS), aiming to maximize the minimum secure energy efficiency through joint optimization of the analog, digital, and true-time delayer (TTD) matrices at the BS, as well as the transmission and reflection coefficients of the STAR-RIS.
% PLS complements conventional cryptography by exploiting inherent wireless channel properties (e.g., interference, fading, noise, and spatial disparity) to enhance security, without the burden of key generation and management \cite{7762075}. 
{\color{black}Existing research on PLS in wideband systems has primarily focused on either far-field \cite{8513857} or near-field scenarios \cite{10879532,10542651,zhangyuchen2023near}. For example,  in \cite{10542651}, the authors proposed a true-time delayer (TTD)-based hybrid beamforming architecture to secure near-field wideband communications. 
Subsequently, TTD-based analog beam-focusing techniques were introduced in \cite{zhangyuchen2023near} to mitigate the impact of the near-field propagation and wideband beam split, thereby boosting the system’s secrecy performance. 
Despite the development of various sophisticated designs to enhance wideband secrecy performance under different system settings, most existing works demonstrate their effectiveness primarily through numerical simulations. The analytical characterization of secrecy performance in wideband PLS remains largely unexplored, particularly in terms of frequency-domain secure behavior. }
%This gap hinders a comprehensive understanding of the fundamental secrecy characteristics of PLS in both far-field and near-field wideband communication systems.}

In practical wireless systems, users are not exclusively located in either the far field or the near field, but often coexist across both regions. 
For example, consider a base station (BS) equipped with $256$ antennas operating at a central frequency of $100$ GHz. The well-known Rayleigh distance, derived based on phase variation, is approximately $98$ meter (m). Therefore, in a typical cellular deployment with a cell radius of $200$ m, users are likely to reside in either the near-field or far-field region of the XL-array \cite{zhang2023mixed,liu2025physical}. The mixed-field communication systems give rise to a unique \emph{energy-spread} effect, which introduces new design opportunities and challenges for the conventional communication system architectures. Specifically, it was shown in \cite{zhang2023mixed} that when the XL-array steers information beams towards the far-field users, the near-field users may suffer strong interference
from far-field-oriented beams, even when the users are located at different angles. This energy-spread effect was further exploited in a simultaneous wireless information and power transfer system \cite{zhang2023swipt}, where far-field-oriented beams are employed to power the surrounding near-field energy harvesting receivers. {\color{black}However, its implications for secrecy performance have so far been examined only in the narrowband regime \cite{liu2025physical}, and a dedicated analysis of how this effect manifests in wideband systems is still lacking in the existing literature.}
\vspace{-11pt}
\subsection{Motivations and Contributions}{\color{black}
More importantly, in the context of wideband PLS, the inherent \emph{energy-spread} effect in mixed-field communications is highly \emph{frequency-dependent} and interacts in a non-trivial manner with the intrinsic frequency-selectivity of wideband channels. This interplay can significantly complicate the secrecy behavior across subcarriers. In particular, the secure regions of mixed-field wideband systems across subcarriers and their fundamental differences from conventional mixed-field (far-field) systems have not yet been theoretically characterized. Furthermore, while AN has been extensively utilized to enhance PLS in conventional narrowband systems \cite{7762075,yunpusecure}, its role and effectiveness in mixed-field wideband PLS scenarios remain unknown.
This raises several critical and unresolved questions: 1) Can AN meaningfully improve secrecy rate in mixed-field wideband scenarios? 2) Can AN leverage the unique near-field beam-focusing effect in such systems? and 3) How should AN adapt to the inherent \emph{frequency-selective} behavior of wideband channels? Despite these important considerations, the study of mixed-field wideband PLS remains largely unexplored, both in terms of fundamental performance analysis and practical transceiver design.}

% Near-field beam-focusing effects provides more design flexibility for beam control in terms of both angle and range domain, thereby significantly improving the performance of numerous wireless applications, such as simultaneous wireless information and power transfer \cite{zhang2023swipt}, physical layer security \cite{yunpusecure}, etc.
{\color{black}
Motivated by the above, we consider in this paper a practical mixed-field wideband PLS system, where a BS equipped with XL-array transmits confidential information to multiple far-field legitimate users in the presence of a near-field eavesdropper. In particular, we address a challenging mixed-field PLS scenario in which the eavesdropper is located closer to the BS than the far-field legitimate users, under which we study the joint beamforming with AN for maximizing the achievable secrecy rate. The main contributions of this paper are summarized as follows.
\begin{itemize}
    \item First, to the best of our knowledge, this work provides the \emph{first} systematic study of PLS in mixed-field wideband communication systems. We theoretically characterize the secure transmission regions across subcarriers and reveal the fundamental differences between the mixed-field wideband PLS and the conventional mixed-field (far-field) PLS systems. Moreover, we formulate an optimization problem to maximize the achievable secrecy rate by jointly optimizing the transmit beamforming with AN for all subcarriers under per-subcarrier power constraints.
    \item Second, to shed useful insights into the new characteristics of mixed-field wideband PLS, we analyze a special case involving a single far-field legitimate user and a near-field eavesdropper. Interestingly, we show that, due to the interplay between energy spread and frequency selectivity, the secure transmission regions vary across subcarriers and are determined by subcarrier-specific \emph{secrecy conditions} derived using the Fresnel integrals. Furthermore, we theoretically demonstrate that incorporating \emph{frequency-selective} AN can significantly enhance the mixed-field wideband PLS by: 1) transforming inherently insecure subcarriers into secure ones, and 2) substantially improving the overall secrecy performance.
    \item Third, for the general system setup, we propose an efficient two-stage hybrid beamforming algorithm based on semidefinite relaxation (SDR) and successive convex approximation (SCA). Building on the analytical insights into the structure of the optimal AN, we further develop a low-complexity beamforming algorithm that removes the need to optimize the AN beamformers on a per-subcarrier basis by directly aligning them with the eavesdropper’s channel, thereby achieving substantial computational complexity reduction while maintaining comparable secrecy performance.
    \item Finally, we present numerical results to demonstrate the unique characteristics of the proposed mixed-field wideband PLS systems and to evaluate the effectiveness of the two proposed algorithms. In particular, it is shown that 1) the secrecy rates across subcarriers are influenced by the subcarrier frequency, indicating a declining trend as the frequency increases; 2) the incorporation of \emph{frequency-selective} AN significantly enhances secrecy rate under different system configurations; and 3) both algorithms significantly outperform the benchmark schemes, with the low-complexity approach offering secrecy performance comparable to that of the more computationally intensive two-stage scheme.
\end{itemize}
}

 \emph{Notations:} Vectors, matrices, and sets are denoted by lowercase boldface letters, uppercase boldface letters, and uppercase calligraphic letters, respectively. The superscripts $(\cdot)^T$ and $(\cdot)^H$ represent the transpose and Hermitian transpose, respectively. The symbol $\odot$ denotes the Hadamard (element-wise) product.
 The notation $\mathcal{CN}(\mu,\sigma^2)$ denotes a complex Gaussian distribution with mean $\mu$ and variance $\sigma^2$. The operator $|\cdot|$ denotes the absolute value when applied to scalars and the cardinality when applied to sets. The Frobenius norm of a matrix is denoted by $\|\cdot\|_F$. For a matrix $\mathbf{X}$, $\mathrm{Tr}(\mathbf{X})$ and $\mathrm{Rank}(\mathbf{X})$ denote its trace and rank, respectively. Moreover, $\mathbf{X} \succeq 0$ indicates that $\mathbf{X}$ is positive semidefinite. $\mathcal{O}(\cdot)$ denotes the standard big-O notation for computational complexity.
 \vspace{-8pt}
\section{System Model and Problem Formulation}\label{Sec:SM}

We consider a mixed near-field and far-field wideband PLS communication system, where a BS equipped with a uniform linear array (ULA) comprising $N=2\Tilde{N}+1$ antennas serves $K$ single-antenna legitimate users, indexed by $\mathcal{K}=\{1,\ldots,K\}$, in the presence of a single-antenna eavesdropper.\footnote{\color{black}The proposed framework can be readily extended to multiple eavesdroppers, where the eavesdropping rate for each legitimate user is determined by the maximum rate achievable among all eavesdroppers. Accordingly, the objective in \eqref{P1:obj} becomes a worst-case secrecy-rate maximization problem, which can be efficiently addressed by introducing auxiliary slack variables to decouple the max operator and applying SCA to handle the resulting non-convex constraints~\cite{liu2025physical}.} The system employs orthogonal frequency division multiplexing
(OFDM) with $M$ subcarriers, indexed by $\mathcal{M}=\{1,\ldots,M\}$ \cite{wang2023beamfocusing}. Specifically, we focus on a practical and challenging communication scenario in which all legitimate users are located in the far-field region of the BS, while the eavesdropper resides in the near-field region and attempts to intercept the transmitted signals at a short distance \cite{liu2025physical}.
Let $B$ denote the total system bandwidth, $f_c$ the central carrier frequency, and $c$ the speed of light. The corresponding central wavelength is given by $\lambda_c = \frac{c}{f_c}$. The total bandwidth is equally divided among the $M$ subcarriers and the frequency of subcarrier $m$ is defined as:
\begin{equation}
    f_m = f_c -\frac{B}{2} + \frac{B(m-1)}{M-1}, \quad \forall m \in \mathcal{M}.
\end{equation}
The antenna spacing of the ULA at the BS is set to half the central wavelength, i.e., $d=\lambda_c/2$.
 \vspace{-11pt}
\subsection{Channel Models}
In this paper, we consider the \emph{effective Rayleigh distance} as the boundary between the near-field and far-field regions, defined as $R_{\rm Ray}(\theta)=\upsilon\sin^2{\theta}\frac{2D^2}{\lambda_c}$, where $\upsilon=0.37$ \cite{cuiwideband}, and $\theta$ represents the user angle, and $D=(N-1)d$ denotes the XL-array aperture. Note that unlike the classic Rayleigh distance obtained based on the maximum allowable phase variations (typically less than $\frac{\pi}{8}$), the effective Rayleigh distance $R_{\rm Ray}(\theta)$ characterizes the boundary beyond which the array gains obtained from the far-field and near-field channel models exhibit a notable difference.
Moreover, both legitimate users and the eavesdropper are modeled using the general multipath channel model, which comprises a line-of-sight (LoS) path and multiple non-LoS (NLoS) components caused by surrounding scatterers \cite{yunpusecure}.\footnote{\color{black}We assume that perfect channel state information (CSI) of both the legitimate users and the eavesdropper is available at the BS to facilitate the fundamental secrecy analysis of mixed-field wideband PLS systems \cite{zhang2023physical}. 
In practice, the CSI of legitimate users can be acquired through existing near-field and far-field beam training or channel estimation techniques \cite{zhang2022fast}. For passive eavesdroppers, obtaining instantaneous CSI is generally challenging due to their non-cooperative nature; however, their presence and location information may be inferred through sensing-assisted approaches \cite{6288501}. RF leakage-based techniques that exploit unintended local oscillator leakage from receiver RF frontend may also provide auxiliary means for passive eavesdropper detection \cite{6288501}. The impact of CSI uncertainty on secrecy performance is further evaluated in Section \ref{Sec:csierror}. Robust beamforming under imperfect CSI can also be  incorporated into the proposed framework by combining SCA with Bernstein-type inequality conversions \cite{9180053}.} The models of near-field and far-field wideband channels are detailed below.

\subsubsection{Near-field Wideband Channel Model}
Based on the spherical wave propagation, the near-field channel between the BS and the eavesdropper on subcarrier $m$ is modeled as 

\begin{align}
    \mathbf{h}^H_{{\rm E}, m} = \sqrt{N}&\beta_{{\rm E},m}\mathbf{b}^H(f_m,\theta_{\rm E}, r_{\rm E})\nn\\
    &+\sqrt{\frac{{N}}{L_{{\rm E}}}}\sum^{L_{{\rm E}}}_{\ell=1}\beta_{{\rm E},m,\ell}\mathbf{b}^H(f_m,\theta_{{\rm E},\ell}, r_{{\rm E},\ell}),
\end{align}
where $\beta_{{\rm E},m}$ and $\beta_{{\rm E},m,\ell}$ denote the complex channel gains of the LoS path and the $\ell$-th NLoS path, respectively, while $L_{{\rm E}}$ is the number of NLoS paths. The angles and ranges for each LoS and NLoS path are denoted by $(\theta_{\rm E}, r_{\rm E})$ and $(\theta_{{\rm E},\ell}, r_{{\rm E},\ell})$, respectively. The near-field channel steering vector $\mathbf{b}^H(f_m,\theta_{\rm E},r_{\rm E})\in \mathbb{C}^{1\times N}$ is given by

\begin{align}\label{Eq:near-field_steering}
    \mathbf{b}^H(f_m,\theta_{\rm E}, r_{\rm E})= 
    \frac{1}{\sqrt{N}}\bigg[e&^{-j\frac{2\pi f_m}{c}(r^{(-\tilde{N})}_{\rm E}-r_{\rm E})},\cdots,\nn\\
    &e^{-j\frac{2\pi f_m}{c}(r^{(\tilde{N})}_{\rm E}-r_{\rm E})}\bigg],
\end{align}
where the distance between the eavesdropper and the $n$-th antenna of the BS is:
\begin{align}\label{Eq:distance}
    r^{(n)}_{\rm E} &= \sqrt{r_{\rm E}^2+n^2d^2-2ndr_{\rm E}\cos{\theta_{\rm E}}}.
\end{align} 
\subsubsection{Far-field Wideband Channel Model}
Under the planar wave assumption, the far-field channel between the BS and legitimate user $k$ on subcarrier $m$ is modeled as 
\begin{align}\label{Eq:far-field_steering}
    \mathbf{h}^H_{{\rm B},m,k} = \sqrt{N}&\beta_{{\rm B},m,k}\mathbf{a}^H(f_m,\phi_{{\rm B},k})\nn\\
    &+\sqrt{\frac{{N}}{L_{{\rm B},k}}}\sum^{L_{{\rm 
    B},k}}_{\ell=1}\beta_{{\rm B},m,k,\ell}\mathbf{a}^H(f_m,\phi_{{\rm B},k,\ell}),
\end{align}
where $\beta_{{\rm B},m,k}$ and $\beta_{{\rm B},m,k,\ell}$ represent the channel gains of the LoS and $\ell$-th NLoS components, respectively, and $L_{{\rm B},k}$ is the number of NLoS paths, while $\phi_{{\rm B},k}$ and $\phi_{{\rm B},k,\ell}$ denote their angles. The corresponding far-field steering vector $\mathbf{a}^H(f_m,\phi_{{\rm B},k})\in \mathbb{C}^{1\times N}$ on subcarrier $m$ is given by
\begin{align}\label{Eq:far-field_steering}
    \mathbf{a}^H(f_m,\phi_{{\rm B},k})= 
    \frac{1}{\sqrt{N}}\bigg[e&^{j\frac{2\pi f_m}{c}\tilde{N}d\cos{\phi_{{\rm B},k}}},\cdots,\nn\\
    &e^{j\frac{2\pi f_m}{c}-\tilde{N}d\cos{\phi_{{\rm B},k}}}\bigg].
\end{align}
It is worth noting that the far-field steering vector in \eqref{Eq:far-field_steering} is a simplified form of the near-field expression in \eqref{Eq:near-field_steering}, which can be obtained by letting the range tend to infinity (i.e., $r \to \infty$).

 \vspace{-12pt}
\subsection{Hybrid Beamforming for Mixed-field Wideband PLS System}
{\color{black}To reduce the number of radio frequency (RF) chains and the associated power consumption required by XL-arrays, we in this paper adopt a hybrid beamforming architecture, wherein the BS is equipped with $N_{\rm RF}$ RF chains $(K\le N_{\rm RF}\ll N)$ to support secure communications for $K$ legitimate users.} However, it is important to note that the conventional hybrid beamforming architecture, which employs the \emph{phase shifter (PS)-based analog beamformer}, is not directly applicable to wideband communication systems \cite{wang2023beamfocusing}. This is because the far-field and near-field channels, as defined in \eqref{Eq:near-field_steering} and \eqref{Eq:far-field_steering}, are \emph{frequency selective} in wideband systems, while the PS-based analog beamforming is \emph{frequency non-adaptive}  across all subcarriers. This mismatch leads to the \emph{beam split} phenomenon, which significantly degrades the achievable rate performance \cite{cuiwideband,10221794}.
To address this issue, TTD-based analog beamformer has been widely adopted to mitigate the beam-split effect by enabling \emph{frequency-dependent} beams, thereby effectively compensating for the near-field beam-split effect \cite{wang2023beamfocusing,cuiwideband,10221794}. Specifically, TTDs introduce a time delay $t$ into the signals, resulting in a frequency-dependent phase shift of $e^{-j{2\pi f_m}t}$ on subcarrier $m$ in the frequency domain \cite{cuiwideband}. In this paper, we consider the \emph{composite} analog beamformer that integrates both $N$ PSs and $N$ TTDs to realize \emph{frequency-adaptive} analog beamforming \cite{cuiwideband}.\footnote{\color{black}Although the fully-connected TTD architecture adopted in this work requires many delay elements and incurs additional hardware cost, it is employed to facilitate the fundamental analysis of mixed-field wideband PLS. Specifically, it compensates for the frequency-dependent propagation delay and eliminates the beam-split effect, thereby allowing us to isolate and characterize the intrinsic interaction between mixed-field energy-spread and wideband frequency-selectivity. More practical TTD architectures, such as sparse or partially connected  structures with fewer delay elements, can further reduce the hardware complexity \cite{wang2023beamfocusing,11482826}. However, these architectures introduce additional coupling among delay allocation, residual beam-split effects, and mixed-field propagation, which requires further investigation and is beyond the scope of this work.} The resulting frequency-dependent analog beamformer on subcarrier $m$, denoted by $\mathbf{F}_{{\rm A},m}\in\mathbb{C}^{N\times N_{\rm RF}}$, is defined as
\begin{equation}
    \mathbf{F}_{{\rm A},m} = \mathbf{F}^{\rm (PS)}_{{\rm A}} \odot\mathbf{F}^{\rm (TTD)}_{{\rm A},m},
\end{equation}
where $\mathbf{F}^{\rm (PS)}_{{\rm A}}\in\mathbb{C}^{N\times N_{\rm RF}}$ and $\mathbf{F}^{\rm (TTD)}_{{\rm A},m}\in\mathbb{C}^{N\times N_{\rm RF}}$ denote the analog beamformer matrices corresponding to the PSs and TTDs, respectively.
\vspace{-8pt}
\subsection{Signal Model} 
Let $\mathbf{x}_m\in\mathbb{C}^{N\times 1}$ denote the transmitted signal vector from the BS to the $K$ legitimate users on subcarrier $m$. This signal consists of both the information-bearing components and AN, and is expressed as
\begin{equation}
    \mathbf{x}_m = \mathbf{F}_{{\rm A},m}\mathbf{F}_{{\rm D},m}\mathbf{s}_m+\mathbf{F}_{{\rm A},m}\mathbf{f}_{{\rm N},m}{z}_m,
\end{equation}
where $\mathbf{F}_{{\rm A},m}\in\mathbb{C}^{N\times N_{\rm RF}}$ is the analog beamformer, $\mathbf{F}_{{\rm D},m}\in\mathbb{C}^{N_{\rm RF}\times K}$ is the digital beamformer for the information-bearing signals, and $\mathbf{f}_{{\rm N},m}\in \mathbb{C}^{N_{\rm RF}\times 1}$ is the digital beamformer for the AN. The vector $\mathbf{s}_m \in \mathbb{C}^{K\times 1}$ denotes the data symbols for the $K$ users on subcarrier $m$. In particular, to ensure secure communication, AN $z_{m}\sim\mathcal{CN}(0,1)$ is injected into the desired signal that is transmitted by the BS to impair the eavesdropper on subcarrier $m$.
Accordingly, the received signal at the legitimate user $k$ on subcarrier $m$ is given by
\begin{align}
    y_{{\rm B},m,k}=&\mathbf{h}^H_{{\rm B},m,k}\mathbf{F}_{{\rm A},m}\mathbf{F}_{{\rm D},m}\mathbf{s}_m\nn\\
    &+\mathbf{h}^H_{{\rm B},m,k}\mathbf{F}_{{\rm A},m}\mathbf{f}_{{\rm N},m}{z}_m+n_{{\rm B},m,k},
\end{align}
where $n_{{\rm B},m,k}\sim \mathcal{CN}(0,\sigma^2_{{\rm B},m,k})$  
is the additive white Gaussian noise (AWGN) at the user $k$ on subcarrier $m$. The corresponding achievable information rate of legitimate user $k$ on subcarrier $m$ is given by \eqref{Eq:bob_rate}, as shown on the top of this page, where $\mathbf{f}_{{\rm D},m,k}$ denotes the $k$-th column of $\mathbf{F}_{{\rm D},m}$.

\begin{figure*}[t]
\begin{align}\label{Eq:bob_rate}
       R^{\rm Sec}_{{\rm B},m,k} =\log_{2}\left( 1+\frac{| \mathbf{h}_{{\rm B},m,k}^H\mathbf{F}_{{\rm A},m}\mathbf{f}_{{\rm D},m,k}| ^2}{\sum^{K}_{i=1,i\neq k}| \mathbf{h}_{{\rm B},m,k}^H\mathbf{F}_{{\rm A},m}\mathbf{f}_{{\rm D},m,i}|^2+| \mathbf{h}_{{\rm B},m,k}^H\mathbf{F}_{{\rm A},m}\mathbf{f}_{{\rm N},m}|^2+\sigma^2_{{\rm B},m,k}}\right),
\end{align}
\vspace{-2pt}
\hrulefill
\vspace{-18pt}
\end{figure*}

The corresponding received signal at the eavesdropper for intercepting
information transmitted to legitimate user $k$ on subcarrier $m$ is given by
\begin{align}
	y_{{\rm E},m,k} &=\mathbf{h}^H_{{\rm E},m}\mathbf{F}_{{\rm A},m}\mathbf{F}_{{\rm D},m}\mathbf{s}_m+\mathbf{h}^H_{{\rm E},m}\mathbf{F}_{{\rm A},m}\mathbf{f}_{{\rm N},m}{z}_m+n_{{\rm E},m},
\end{align}
where $n_{{\rm E},m}\sim \mathcal{CN}(0,\sigma^2_{{\rm E},m})$  
denotes the AWGN at the eavesdropper. Specifically, we assume that the eavesdropper is capable of canceling all
multiuser interference before decoding the desired information \cite{yunpusecure}.
% \footnote{The proposed algorithm can be readily extended to scenarios that account for the interference from legitimate users to the eavesdropper in \eqref{Eq:informationLeak}, by applying a Taylor approximation of function $D_{m,2}$ with respect to all beamforming matrices $\mathbf{W}_{{\rm D},m}$ \cite{9525400,10879532}.} 
Thus, the channel capacity between the BS and the eavesdropper for wiretapping legitimate user $k$ on subcarrier $m$ is given by
\begin{align}\label{Eq:informationLeak}
	C_{{\rm E},m}=\log_{2} \left(1+\frac{|\mathbf{h}_{{\rm E},m}^H\mathbf{F}_{{\rm A},m}\mathbf{f}_{{\rm D},m,k}|^2}{|\mathbf{h}_{{\rm E},m}^H\mathbf{F}_{{\rm A},m}\mathbf{f
    }_{{\rm N},m}|^2+\sigma^2_{{\rm E},m}} \right).
\end{align}
The achievable secrecy rate of the legitimate user $k$ on subcarrier $m$ is given by
$R^{\rm Sec}_{{\rm B},m,k}=\left[R_{{\rm B},m,k}-C_{{\rm E},m}\right]^{+}$, where $[x]^{+}\triangleq\max\{x,0\}$. 
\vspace{-8pt}
\subsection{Problem Formulation}
In this paper,
we aim to maximize the achievable secrecy rate of the considered mixed near-field and far-field wideband communication system by jointly designing the analog and digital beamformers, namely, $\mathbf{F}_{{\rm A},m}$, $\mathbf{F}_{{\rm D},m}$ and $\mathbf{f}_{{\rm N},m}$. The optimization problem is formulated as\footnote{\color{black}
Problem (P1) can be readily extended to other scenarios by adapting the near-field and far-field channel models. When both the legitimate users and the eavesdropper are in the near field, secure transmission is generally achievable unless the eavesdropper is closer to the BS in range. When both are in the far field, secrecy may be compromised if they share the same angular direction. When the legitimate users are in the near field and the eavesdropper is in the far field, secure transmission is typically achievable due to the eavesdropper’s higher path loss and lack of near-field beam-focusing gain.}
\begin{subequations}
	\begin{align}
		({\bf P1}):\max_{\substack{\{\mathbf{F}_{{\rm A},m}, \mathbf{F}_{{\rm D},m},\\
        \mathbf{f}_{{\rm N},m}\} }}  &\frac{1}{M}\sum^{M}_{m=1}\sum^{K}_{k=1}R^{\rm Sec}_{{\rm B},m,k}\label{P1:obj}
		\\
		\text{s.t.}~~~~~
		&	\|\mathbf{F}_{{\rm A},m}\mathbf{F}_{{\rm D},m}\|^2_F+\|\mathbf{F}_{{\rm A},m}\mathbf{f}_{{\rm N},m}\|^2\le P,\label{P1:pow_cons}\\
        & |\mathbf{F}_{{\rm A},m}(i,j)|=1,\forall{(i,j)}\in\mathcal{F},\label{P1:ana_norm}
	\end{align}
 \end{subequations}
where $P$ denotes the maximum available transmit power  per subcarrier, and \eqref{P1:ana_norm} imposes the unit-modulus constraint on the analog beamformer for subcarrier $m$, with the set $\mathcal{F}$ containing the indices of non-zero entries in $\mathbf{F}_{{\rm A},m}$.

Problem (P1) is generally difficult to solve due to the following reasons: 1) the objective function is non-convex; 2) the optimization variables are highly coupled across different subcarriers; and 3) intrinsically non-convex unit-modulus constraints. To address these challenges and provide useful insights into the design of mixed-field wideband PLS, we first study a special case in Section \ref{Sec:Special_case}, and then develop two efficient algorithms to solve problem (P1) in Section \ref{Sec:general}.

\vspace{-8pt}
\section{Secrecy Performance Analysis for Mixed-Field Wideband PLS} \label{Sec:Special_case}
To shed important insights into the new characteristics of mixed near-field and far-field wideband PLS, we consider in this section a special case involving a single legitimate user ($K=1$). Interestingly, we reveal that the secure transmission conditions of mixed-field wideband systems are different over subcarriers due to the wideband \emph{frequency selectivity}, which is in sharp contrast to the conventional mixed-field narrowband PLS, for which the secure transmission condition is determined solely by the central carrier frequency \cite{liu2025physical}. Furthermore, we theoretically demonstrate that incorporating \emph{frequency-selective} AN to mixed-field wideband PLS brings two key advantages: 1) it can transform insecure subcarriers into secure ones, and 2) it significantly enhances the overall secrecy performance.

For notational simplicity, we drop the user index in the following analysis. To theoretically reveal the unique characteristics inherent in the mixed-field wideband PLS, we analyze the secrecy rate on a per-subcarrier basis. Accordingly, the achievable secrecy rate on subcarrier $m$ can be expressed as
% \footnote{For analytical simplicity, the secrecy rate in \eqref{Eq:SC_rate_expression} is obtained by assuming that the legitimate user and the eavesdropper experience the same noise power on each subcarrier, i.e., $\sigma^2_{{\rm E},m}=\sigma^2_{{\rm E},m}=\sigma^2_{m}$. For the case where the noise powers differ, this expression still holds after normalizing the
% noise power by the corresponding channel gains, i.e., $\beta_{{\rm B},m}$ and $\beta_{{\rm E},m}$ \cite{yunpusecure}.}
\begin{align}\label{Eq:SC_rate_expression}
     R^{{\rm Sec}}_{{\rm B},m}
    = &\log_{2}\left( 1+\frac{| \mathbf{h}_{{\rm B},m}^H\mathbf{F}_{{\rm A},m}\mathbf{f}_{{\rm D}}| ^2}{| \mathbf{h}_{{\rm B},m}^H\mathbf{F}_{{\rm A},m}\mathbf{f}_{{\rm N},m}| ^2+\sigma^2_{m}}\right)\nn\\
     &-\log_{2} \left(1+\frac{|\mathbf{h}_{{\rm E},m}^H\mathbf{F}_{{\rm A},m}\mathbf{f}_{{\rm D}}|^2}{|\mathbf{h}_{{\rm E},m}^H\mathbf{F}_{{\rm A},m}\mathbf{f}_{{\rm N},m}|^2+\sigma^2_{m}} \right).
\end{align}

Clearly, the secrecy rate expression in \eqref{Eq:SC_rate_expression} is still in a complicated form, making it difficult to analyze. To tackle this difficulty, we consider a LoS-only propagation scenario, which is widely adopted in high-frequency bands \cite{liu2025physical,wang2023beamfocusing}.\footnote{\color{black}The LoS-dominant assumption is adopted to extract the deterministic geometric characteristics underlying mixed-field wideband PLS. Although NLoS components introduce additional channel fluctuations, the derived analytical framework captures the fundamental secrecy behavior governed by the mixed-field propagation geometry. Extending the closed-form secrecy  characterization to severely blocked LoS or rich-scattering environments requires a stochastic correlation analysis, which is beyond the scope of this work.} Moreover, a low-complexity hybrid beamforming design is considered \cite{zhang2023swipt}, for which the analog beamformer is devised based on the maximal ratio transmission (MRT) to simultaneously steer beams toward both the legitimate user and the eavesdropper; while the power allocation is optimized in the digital beamforming stage.\footnote{It is worth noting that while more advanced digital beamformers such as zero-forcing (ZF) or minimum-squared error (MMSE) may offer enhanced performance, they significantly complicate the analysis and may even render it intractable. To enable analytical insights, we focus on power allocation-based digital beamforming in this section. The general optimization of digital beamformers will be carried out in Section \ref{Sec:general}.} Under this setup, the composite analog beamformers are given by 
\begin{equation}
    \mathbf{F}_{{\rm A},m}=\sqrt{N}\left[\mathbf{a}(f_m,\phi_{{\rm B}}),\mathbf{b}(f_m,\theta_{\rm E}, r_{\rm E})\right],
\end{equation}
where
\begin{align}
    \mathbf{F}^{\rm (PS)}_{{\rm A}} &= \sqrt{N}[\mathbf{a}(f_c,\phi_{{\rm B}}),\mathbf{b}(f_c,\theta_{\rm E}, r_{\rm E})],\\ \mathbf{F}^{\rm (TTD)}_{{\rm A},m} &= [e^{-j{2\pi f_m}\mathbf{t}_{\rm B}},e^{-j{2\pi f_m}\mathbf{t}_{\rm E}}],
\end{align}
with 
\begin{align}
    \mathbf{t}_{\rm B} &= \frac{\angle{\mathbf{a}(f_c,\phi_{{\rm B}})}-\angle{\mathbf{a}(f_m,\phi_{{\rm B}})}}{2\pi f_m},\\
    \mathbf{t}_{\rm E} &= \frac{\angle{\mathbf{b}(f_c,\theta_{\rm E}, r_{\rm E})}-\angle{\mathbf{b}(f_m,\theta_{\rm E}, r_{\rm E})}}{2\pi f_m}. 
\end{align}
% with $\tilde{\mathbf{a}}(f_m,\phi_{{\rm B}}) =e^{-j(\angle{\mathbf{a}(f_m,\phi_{{\rm B}})}-\angle{\mathbf{a}(f_c,\phi_{{\rm B}})})}$ $\tilde{\mathbf{b}}(f_m,\theta_{\rm E}, r_{\rm E})=e^{-j(\angle{\mathbf{b}(f_m,\theta_{\rm E}, r_{\rm E})}-\angle{\mathbf{b}(f_c,\theta_{\rm E}, r_{\rm E})})}$.
This leads to the composite beamforming vectors  $\mathbf{F}_{{\rm A},m}\mathbf{f}_{{\rm D},m}=\sqrt{P_{{\rm B},m}}\mathbf{a}(f_m,\phi_{{\rm B}})$ and $\mathbf{F}_{{\rm A},m}\mathbf{f}_{{\rm N},m}=\sqrt{P_{{\rm E},m}}\mathbf{b}(f_m,\theta_{\rm E}, r_{\rm E})$, where $P_{{\rm B},m}$ and $P_{{\rm E},m}$ represent the powers allocated to the information-bearing signal and AN, respectively, with $P_{{\rm B},m}+P_{{\rm E},m}=P$.\footnote{
% In this paper, to obtain analytical insights into mixed-field wideband PLS, we consider the design of the analog beamformer using composite analog beamformers that integrate TDDs and PSs. This architecture enables the generation of \emph{frequency-selective} beams directed towards the users without introducing beam split issues \cite{cuiwideband}. 
In practice, due to hardware limitations, the time delays introduced by TTDs cannot be arbitrarily configured, which may result in performance degradation when fixed delays are used. However, the analytical insights and the proposed framework in this study can be readily extended to the case with the maximum allowable time delay taken into account, since the framework is built upon the effective channels obtained from any analog beamformer design, including non-ideal ones \cite{wang2023beamfocusing}.}
% Moreover, the optimization of TTD configurations under practical constraints has been investigated in the existing literature \cite{10542651,wang2023beamfocusing,10723247}.} 
Based on the above, the achievable secrecy rate on subcarrier $m$ is given by \eqref{Eq:SC_rate_expression1}, as shown on the bottom of this page. It is readily observed that the secrecy rate is significantly influenced by the \emph{cross-correlation} terms, i.e., $|\mathbf{b}^H(f_m,\theta_{\rm E}, r_{\rm E})\mathbf{a}(f_m,\phi_{{\rm B}})|$ and $|\mathbf{a}^H(f_m,\phi_{{\rm B}})\mathbf{b}(f_m,\theta_{\rm E}, r_{\rm E})|$. To analytically characterize these terms and examine the impact of power allocation on mixed-field wideband PLS, we first present their mathematical formulation in Definition \ref{Def:cross_corre}, and then attain a tractable closed-form approximation based on the Fresnel integrals.
  \begin{figure}[t]
\begin{minipage}{.24\textwidth}
	\centering
\includegraphics[width=1\columnwidth]{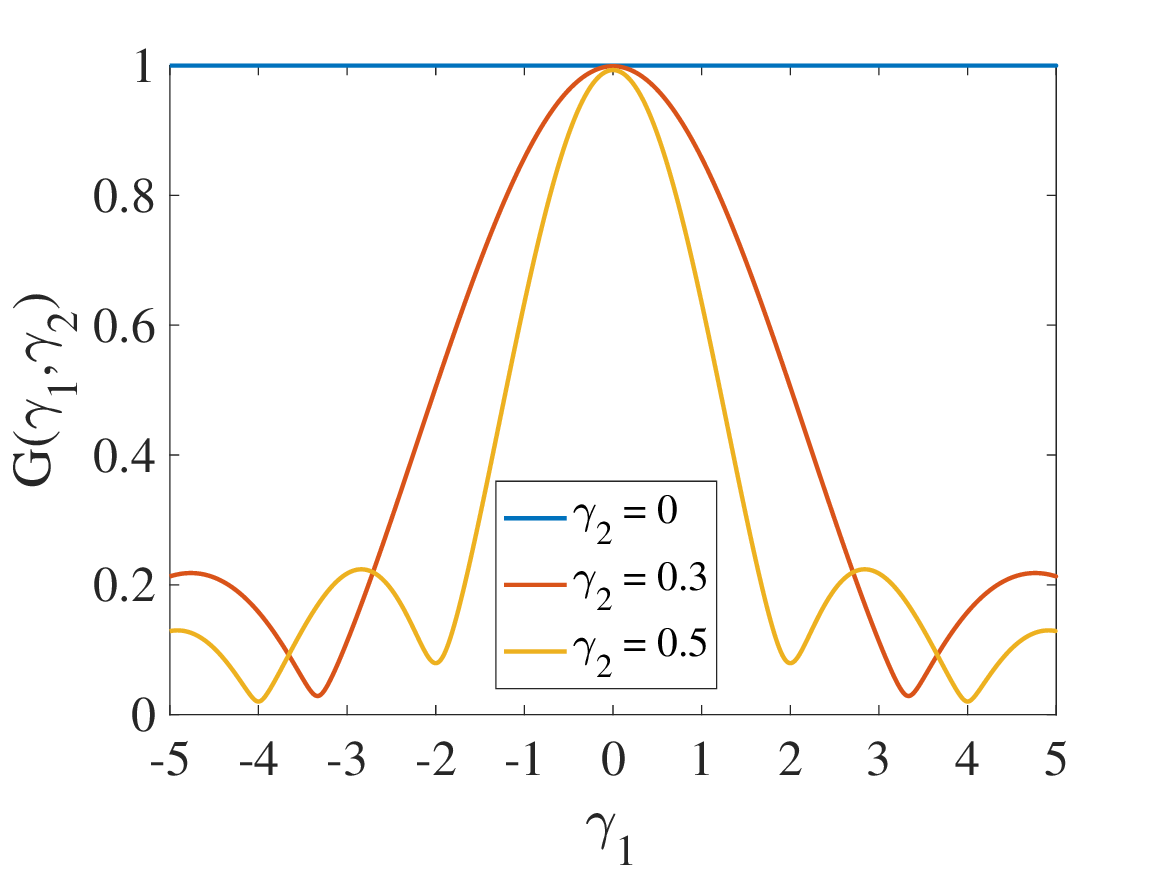}
	\caption{{$G(\gamma_1,\gamma_2)$ versus $\gamma_2$ for fixed $|\gamma_1|$.}\label{Fig:corre}} 
    \end{minipage}	
    \hfill
    \begin{minipage}{.24\textwidth}
	\centering
\includegraphics[width=1\columnwidth]{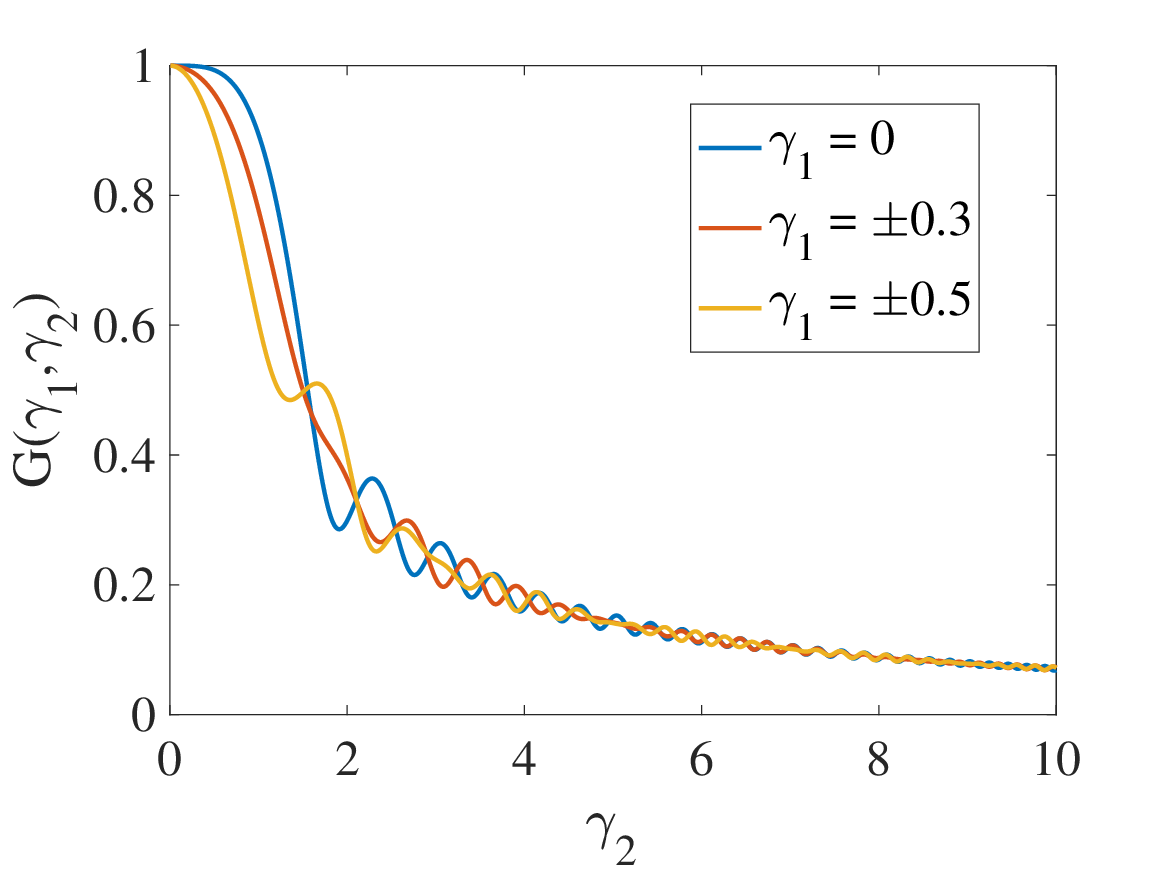}
	\caption{{$G(\gamma_1,\gamma_2)$ versus $\gamma_1$ for fixed $\gamma_2$.}\label{Fig:corre1}} 
    \end{minipage}	
    \vspace{-18pt}
\end{figure}
\begin{figure*}[b]
\vspace{-12pt}
\hrulefill
\begin{align}\label{Eq:SC_rate_expression1}
     R^{{\rm Sec}}_{{\rm B},m}
    = &\log_{2}\!\left(\! 1\!+\!\frac{{P_{{\rm B},m}}N|\beta_{{\rm B},m}|^2}{{P_{{\rm E},m}}N|\beta_{{\rm B},m}|^2|\mathbf{a}^H(f_m,\phi_{{\rm B}})\mathbf{b}(f_m,\theta_{\rm E}, r_{\rm E})| ^2+\sigma^2_{m}}\!\right)\!-\!\log_{2} \!\left(\!1\!+\!\frac{{P_{{\rm B},m}}N|\beta_{{\rm E},m}|^2|\mathbf{b}^H(f_m,\theta_{\rm E}, r_{\rm E})\mathbf{a}(f_m,\phi_{{\rm B}})| ^2}{{P_{{\rm E},m}}N|\beta_{{\rm E},m}|^2+\sigma^2_{m}} \!\right).
\end{align}
\end{figure*}

\begin{definition}\label{Def:cross_corre}
    \emph{The cross-correlation of a near-field steering vector and a far-field steering vector on subcarrier $m$ is defined as:
    \begin{align}
        &\rho(f_m,\theta_{\rm E}, r_{\rm E},\phi_{\rm B}) \nn\\
        &\triangleq |\mathbf{b}^H(f_m,\theta_{\rm E}, r_{\rm E})\mathbf{a}(f_m,\phi_{{\rm B}})|=|\mathbf{a}^H(f_m,\phi_{{\rm B}})\mathbf{b}(f_m,\theta_{\rm E}, r_{\rm E})| \label{Eq:NF_corre_real}\\
        & \overset{(a)}{\approx}\frac{1}{N}\left|\sum_{n=-\tilde{N}}^{\tilde{N}}e^{j\frac{2\pi f_m}{c}\left({nd}(\cos{\phi_{\rm B}}-\cos{\theta_{\rm E}}) + n^2d^2\frac{\sin^2{\theta_{\rm E}}}{2r_{\rm E}}\right)}\right|,\label{Eq:NF_corre}
    \end{align}
    where $(a)$ is obtained by applying a second-order Taylor expansion to the distance expression in \eqref{Eq:distance}, i.e., $\sqrt{1+x}=1+\frac{1}{2}x-\frac{1}{8}x^2+\mathcal{O}(x^3)$, resulting in $r^{(n)}_{\rm E}\approx r_{\rm E}-nd\cos{\theta_{\rm E}}+\frac{n^2d^2\sin^2{\theta_{\rm E}}}{2r_{\rm E}}$ \cite{zhang2023mixed}. 
    {\color{black} This approximation is validated in \cite{7942128} to be sufficiently accurate when the BS-user distance is larger than the Fresnel distance $0.5\sqrt{\frac{D^3}{\lambda_c}}$, which is much smaller than the effective Rayleigh distance considered in this paper.}}
\end{definition}
Next, we provide a closed-form approximation for the cross-correlation in the following lemma using the Fresnel integrals.

\begin{lemma}\label{Le:fresnel}
\emph{The cross-correlation in \eqref{Eq:NF_corre} can be approximated as 
\begin{align}\label{Eq:NF_corre_approx}
    \rho&(f_m,\theta_{\rm E}, r_{\rm E},\phi_{\rm B}) \approx |G(\gamma_1,\gamma_2)| = \left|\frac{\hat{C}(\gamma_1,\gamma_2)+j\hat{S}(\gamma_1,\gamma_2)}{2\gamma_2}\right|,
\end{align}
where $\hat{C}(\gamma_1,\gamma_2) = {C}(\gamma_1+\gamma_2)-{C}(\gamma_1-\gamma_2)$ and $\hat{S}(\gamma_1,\gamma_2) = {S}(\gamma_1+\gamma_2)-{S}(\gamma_1-\gamma_2)$, with the Fresnel integrals defined as $C(x)=\int^{x}_{0}\cos(\frac{\pi}{2}t^2)\text{d}t$ and $S(x)=\int^{x}_{0}\sin(\frac{\pi}{2}t^2)\text{d}t$; and 
\begin{align}
  \gamma_1 &=\frac{f_m}{f_c}(\cos{\phi_{\rm B}}-\cos{\theta_{\rm E}})\sqrt{\frac{r_{\rm E}}{\frac{f_m}{f_c}d\sin^2{\theta_{\rm E}}}},\\
    \gamma_2 &= \frac{N}{2}\sqrt{\frac{\frac{f_m}{f_c}d\sin^2{\theta_{\rm E}}}{r_{\rm E}}}.
\end{align}}
\end{lemma}
\begin{proof}
    Please refer to Appendix \ref{App0}.
\end{proof}
Lemma \ref{Le:fresnel} presents a tractable closed-form expression for the cross-correlation function, which facilitates analytical characterization via the properties of the function $G(\cdot)$. Specifically, $G(\cdot)$ is characterized by two key parameters, $\gamma_1$ and $\gamma_2$. The function is symmetric with respect to $\gamma_1$, generally decreases with $|\gamma_1|$ (see Fig. \ref{Fig:corre}), and exhibits a decreasing trend with increasing $\gamma_2$ (see Fig. \ref{Fig:corre1}) \cite{zhang2023mixed}.

\begin{remark}[Correlation differences between mixed-field wideband and narrowband PLS systems]\label{Re:remark1}
    \emph{Based on the properties of $G(\cdot)$, several key differences distinguish mixed-field wideband PLS systems from their conventional narrowband counterparts:
    \begin{itemize}
        \item First, the cross-correlation expression in \eqref{Eq:NF_corre_approx} serves as a generalized form of the mixed-field narrowband cross-correlation, which reduces to the narrowband form when $f_m = f_c$ \cite{zhang2023mixed}. Mathematically, its cross-correlation is given by $\rho(f_c,\theta_{\rm E}, r_{\rm E},\phi_{\rm B}) \approx |G(\tilde{\gamma_1},\tilde{\gamma_2})|$, where 
  $\tilde{\gamma_1} =(\cos{\phi_{\rm B}}-\cos{\theta_{\rm E}})\sqrt{\frac{r_{\rm E}}{d\sin^2{\theta_{\rm E}}}}$ and 
    $\tilde{\gamma_2} = \frac{N}{2}\sqrt{\frac{d\sin^2{\theta_{\rm E}}}{r_{\rm E}}}$.
        \item  Second, in contrast to narrowband systems where the cross-correlation $\rho(f_c,\theta_{\rm E}, r_{\rm E},\phi_{\rm B})$ is frequency-independent, the cross-correlation in mixed-field wideband systems is significantly influenced by the subcarrier frequency $f_m$. 
        Specifically, it can be easily verified that both $\gamma_1$ and $\gamma_2$ decrease with $f_m$, indicating that the cross-correlation generally decreases with the frequency band. This is intuitively expected, as a higher frequency band with a shorter wavelength leads to a more pronounced effective Rayleigh distance, thereby enhancing the near-field effect, resulting in much higher spatial resolution and consequently lower correlation.
        \item Finally, the effect of the angular difference between the legitimate user and the eavesdropper, i.e., $\cos{\theta_{\rm E}}-\cos{\phi_{\rm B}}$, remains consistent with the narrowband case. It exhibits symmetry in the angular domain and applies uniformly across all subcarriers.
    \end{itemize}
    }
\end{remark}

Next, to reveal the distinct characteristics of mixed-field wideband PLS systems in comparison to their narrowband counterparts, we characterize the corresponding secrecy conditions in the absence of AN.
\vspace{-8pt}
\subsection{Secrecy Conditions for Narrowband and Wideband Systems without AN}

First, we present the secrecy conditions for mixed-field narrowband and wideband systems in the following lemmas.
\begin{lemma}[Narrowband secrecy condition]\label{Le:Narrowband}
	\emph{For mixed-field narrowband systems, the data transmission to legitimate users is secure only when
	         \begin{equation}\label{Eq:NarrowbandSC}
        \boxed{\rho(f_c,\theta_{\rm E}, r_{\rm E},\phi_{\rm B}) \le \frac{r_{\rm E}}{r_{\rm B}}.} 
    \end{equation}}
\end{lemma}
{\begin{proof}\color{black}
First, for the system without AN, the secrecy rate in \eqref{Eq:SC_rate_expression1} can be rewritten as
\begin{align}
    R^{{\rm Sec}}_{{\rm B},m}
    \!=\! \log_{2} \left(\frac{P_{{\rm B},m}| \beta_{{\rm B},m}|^2+\sigma^2_{m}}{P_{{\rm B},m}| \beta_{{\rm E},m}| ^2\rho^2(f_c,\theta_{\rm E}, r_{\rm E},\phi_{\rm B})+\sigma^2_{m} } \right).
\end{align}
To establish the narrowband secrecy condition, we require $ R^{{\rm Sec}}_{{\rm B},m} \ge 0$. Accordingly, the secrecy condition can be expressed as $\rho(f_c,\theta_{\rm E}, r_{\rm E},\phi_{\rm B}) \le \frac{r_{\rm E}}{r_{\rm B}}$,
completing the proof.
\end{proof}}

\begin{lemma}[Wideband secrecy condition]\label{Le:Wideband}
	\emph{For mixed-field wideband systems, the data transmission to legitimate users on subcarrier $m$ is secure only when
		         \begin{equation}\label{Eq:Wideband}
        \boxed{ \rho(f_m,\theta_{\rm E}, r_{\rm E},\phi_{\rm B})  \le \frac{r_{\rm E}}{r_{\rm B}}.} 
    \end{equation}}
\end{lemma}
\begin{proof}
  \color{black}The proof is similar to that of Lemma \ref{Le:Narrowband}, for which the secrecy condition on each subcarrier is obtained by replacing the corresponding frequency in \eqref{Eq:NarrowbandSC}.
\end{proof}
{\color{black}
\begin{remark}[Comparison between narrowband and wideband secrecy conditions]\label{Re:two_conditions}
    \emph{Lemmas \ref{Le:Narrowband} and \ref{Le:Wideband} offer several important insights into mixed-field PLS design:
    \begin{itemize}
        \item The mixed-field narrowband secrecy condition in \eqref{Eq:NarrowbandSC} is \emph{frequency-independent} and is determined solely by the near-field eavesdropper's angle and range, the far-field legitimate user's angle, and the central subcarrier frequency. This implies that once these parameters are fixed, the narrowband secrecy condition remains unchanged.
        \item In contrast, the mixed-field wideband secrecy condition in \eqref{Eq:Wideband} is \emph{frequency-dependent}, varying significantly with the subcarrier frequency. This behavior differs fundamentally from conventional narrowband PLS systems (near-field or mixed-field), where the secrecy condition is determined solely by the central carrier frequency \cite{liu2025physical}.
        \item More importantly, the \emph{frequency-selective} nature of the wideband cross-correlation leads to a new phenomenon:  the data transmissions on certain subcarriers may satisfy the secrecy condition, while others may not. This reveals new opportunities for subcarrier-level secure resource allocation in mixed-field wideband PLS systems.
    \end{itemize}
    }
\end{remark}}

% However, it is worth noting that the expressions for these secure conditions remain complex, making it difficult to oanalyze their impact and identify key influencing parameters. To address this issue, we in the next derive their closed-form and more tractable expressions using the Fresnel integrals.

% \begin{lemma}\label{The:lemma1}
%     \emph{The correlation $ \rho(f_m,f_n,\theta,r)$ in \eqref{Eq:NN_corre} can be approximated as:
%     \begin{align}\label{Eq:rot_corre}
%          \rho(f_m,f_n,\theta,r) \approx G(\beta_1,\beta_2) = \left|\frac{\widehat{C}(\beta_1,\beta_2)+j\widehat{S}(\beta_1,\beta_2)}{2\beta_2}\right|,
%     \end{align}
%     where 
%      \begin{align}
% \beta_1&= \frac{(f_m-f_n)d\cos{\theta}}{c}\sqrt{\frac{2cr}{(f_m-f_n)d^2\sin^2{\theta}}}\nn\\
% \beta_2&={N} \sqrt{\frac{(f_m-f_n)d^2\sin^2{\theta}}{2cr}}.
%     \end{align}
%      Moreover, $\widehat{C}(\beta_1,\beta_2) = {C}(\beta_1+\beta_2)-{C}(\beta_1-\beta_2)$ and $\widehat{S}(\beta_1,\beta_2) = {S}(\beta_1+\beta_2)-{S}(\beta_1-\beta_2)$, with the Fresnel integrals defined as $C(x)=\int^{x}_{0}\cos(\frac{\pi}{2}t^2)\text{d}t$ and $S(x)=\int^{x}_{0}\sin(\frac{\pi}{2}t^2)\text{d}t$.
%     }
% \end{lemma}
Next, we delve into characterizing the effects of the key parameters on the secrecy performance of mixed-field wideband PLS systems.

\begin{proposition}[Secure subcarrier index set]\label{Le:secure_sub}
    \emph{Given the locations of a near-field eavesdropper $(\theta_{\rm E}, r_{\rm E})$ and a far-field legitimate user $(\phi_{\rm B}, r_{\rm B})$, the set of subcarriers that satisfy the secure transmission condition is given by:
    \begin{equation}
    \!\!\mathcal{A}(\theta_{\rm E}, r_{\rm E},\phi_{\rm B}, r_{\rm B}) \!=\!\begin{cases}\!
\{ m \in \mathcal{M} \mid \hat{m} \le m \le M \}, & \text{if } \hat{m} < M \\
\varnothing, & \text{if } \hat{m} \ge M,
\end{cases}
    \end{equation}
    where $\hat{m} = \lceil m^*\rceil$, and $m^*$ is the unique solution to $\rho(f_m^*,\theta_{\rm E}, r_{\rm E},\phi_{\rm B})=\frac{r_{\rm E}}{r_{\rm B}}$.
    }
\end{proposition}
\begin{proof}
    It follows from \eqref{Eq:NF_corre_approx} that the cross-correlation $\rho(f_m,\theta_{\rm E}, r_{\rm E},\phi_{\rm B})$ generally decreases with increasing subcarrier index $m$ \cite{zhang2023mixed}. This implies that lower-frequency subcarriers exhibit higher correlation values, while higher-frequency subcarriers yield lower correlations. To determine the set of secure subcarriers, it is sufficient to identify the smallest subcarrier index $\hat{m}$ such that the secrecy condition in \eqref{Eq:Wideband} is satisfied. Denote by $m^*$ the unique solution to $\rho(f_{m^*},\theta_{\rm E}, r_{\rm E},\phi_{\rm B}) = \frac{r_{\rm E}}{r_{\rm B}}$. Then, the secure subcarrier index support corresponds to $\{ m \in \mathcal{M} \mid \hat{m} \le m \le M\}$, where $\hat{m} = \lceil m^* \rceil$.
\end{proof}
 \begin{figure*}[t!]
\begin{minipage}[t]{.24\textwidth}
    \vspace{0pt}
    \centering
    \includegraphics[width=\linewidth]{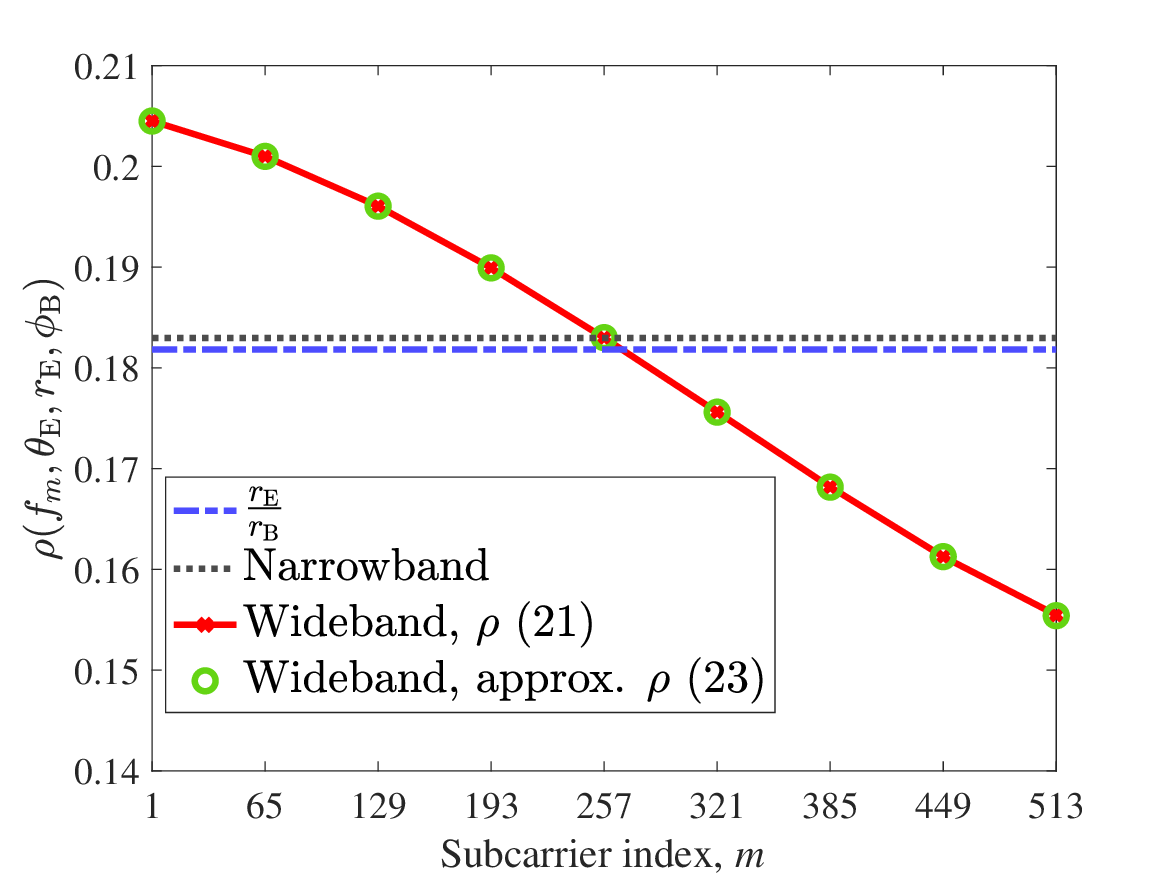}
    \caption{\color{black}Secrecy condition versus subcarrier index with
    $|\theta_{\rm E}-\phi_{\rm B}|=0.01\pi$,
    $r_{\rm E}=0.2R_{\rm Ray}$ and $r_{\rm B}=1.1R_{\rm Ray}$.}
    \label{Fig:sc_frequency}
\end{minipage}
\hfill
\begin{minipage}[t]{.24\textwidth}
    \vspace{0pt}
    \centering
    \includegraphics[width=\linewidth]{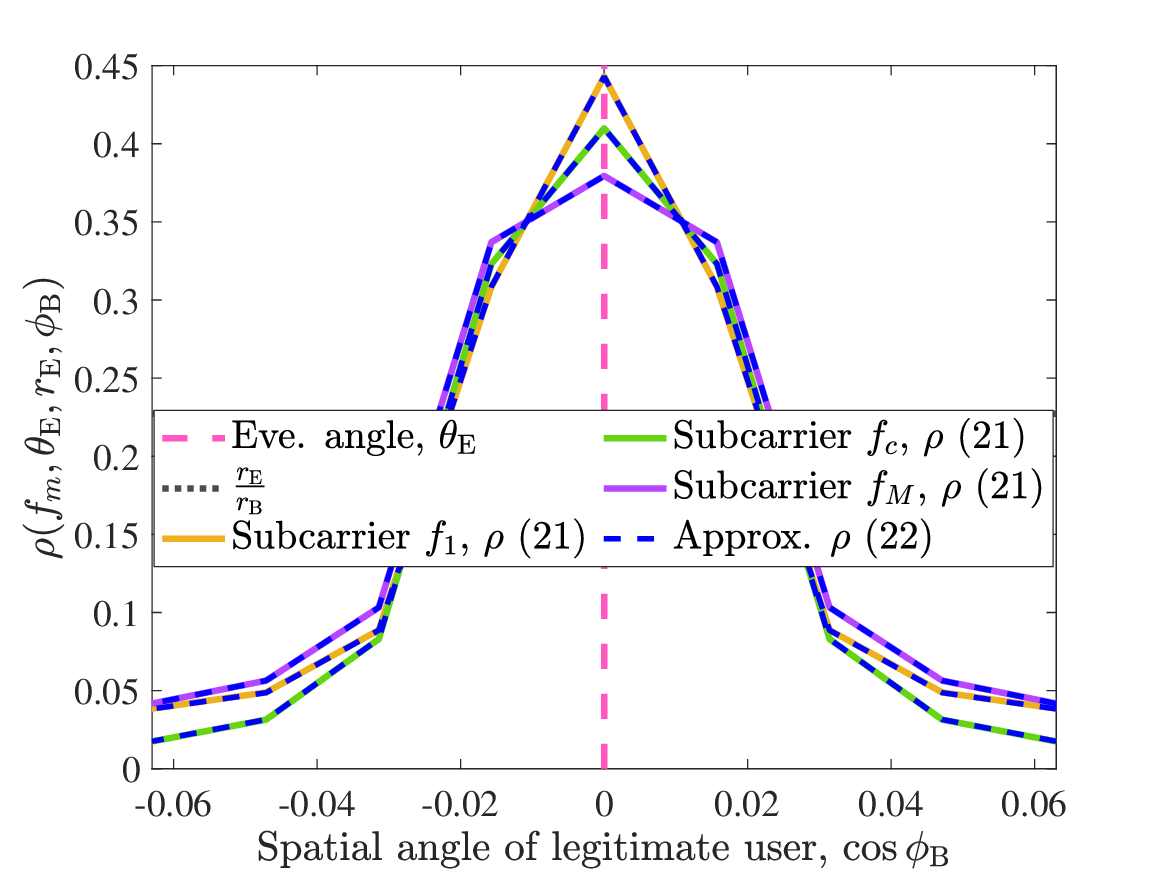}
    \caption{Secrecy condition versus spatial angle of legitimate user with
    $\theta_{\rm E}=0.5\pi$, $r_{\rm E}=0.2R_{\rm Ray}$ and
    $r_{\rm B}=1.1R_{\rm Ray}$.}
    \label{Fig:sc_angle}
\end{minipage}
\hfill
\begin{minipage}[t]{.24\textwidth}
    \vspace{0pt}
    \centering
    \includegraphics[width=\linewidth]{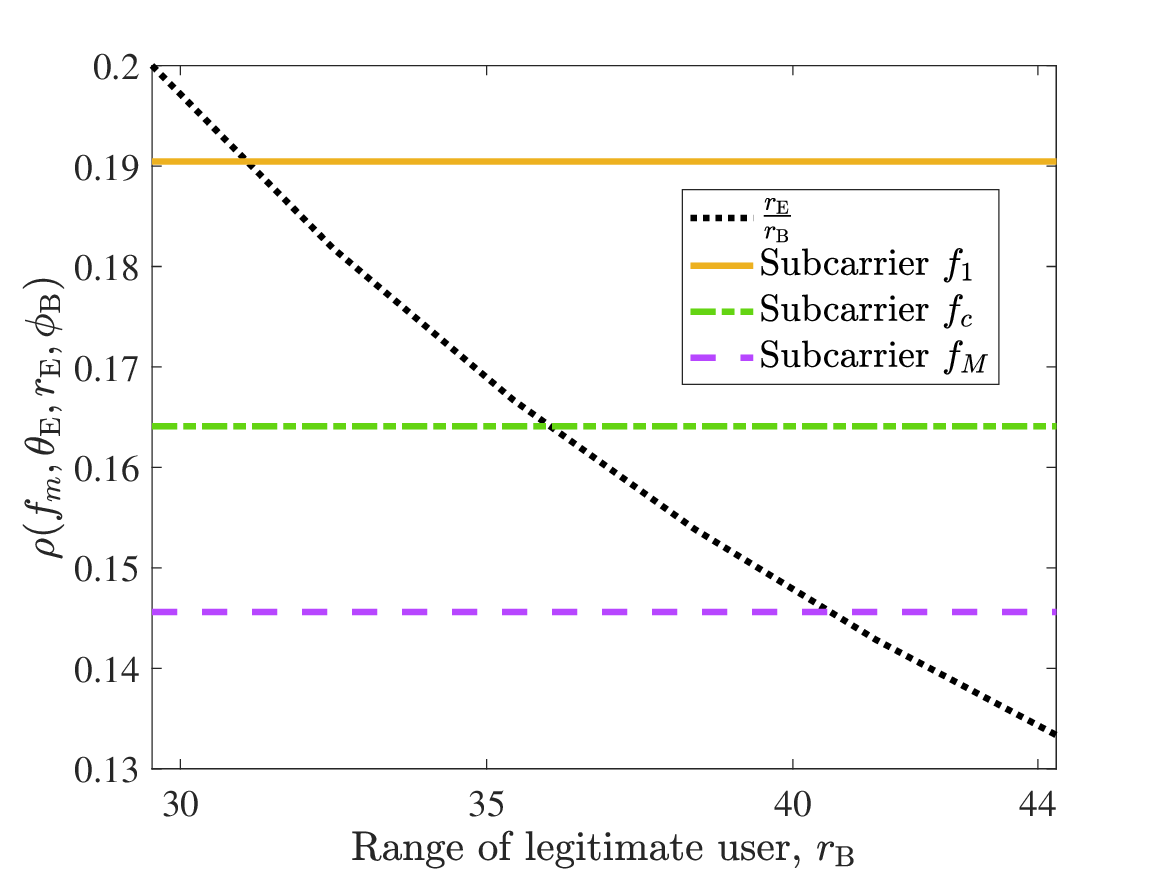}
    \caption{Secrecy condition versus range of legitimate user with
    $|\theta_{\rm E}-\phi_{\rm B}|=0.01\pi$ and
    $r_{\rm E}=0.2R_{\rm Ray}$.}
    \label{Fig:sc_range}
\end{minipage}
\hfill
\begin{minipage}[t]{.24\textwidth}
    \vspace{0pt}
    \centering
    \includegraphics[width=\linewidth]{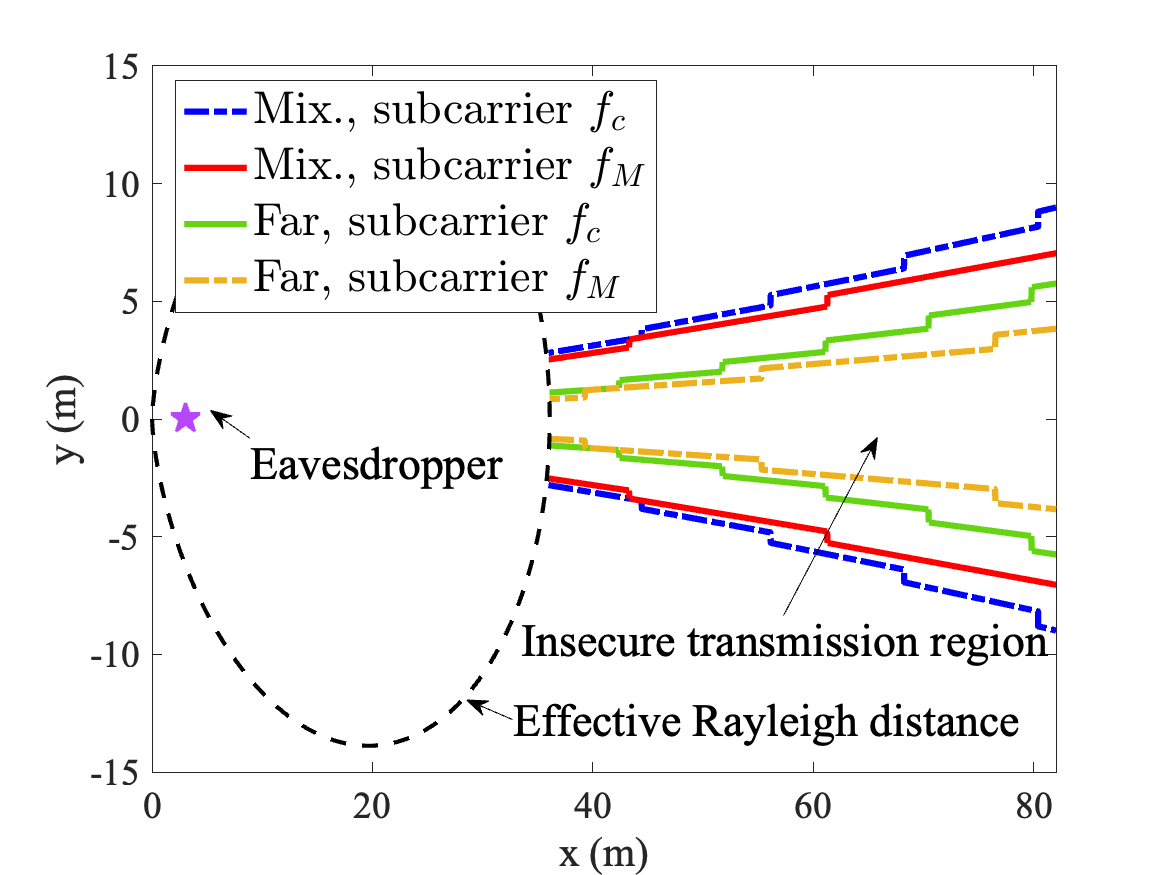}
    \caption{Insecure transmission region of legitimate user on different
    subcarriers with $\theta_{\rm E}=0.5\pi$ and
    $r_{\rm E}=0.2R_{\rm Ray}$.}
    \label{Fig:sc_2dregion}
\end{minipage}
\vspace{-16pt}
\end{figure*}
{\color{black}
\begin{example}
\emph{In Fig. \ref{Fig:sc_frequency}, we plot the secrecy condition versus the subcarrier index and validate the accuracy of the cross-correlation approximation derived in Lemma \ref{Le:fresnel}, with parameters set as $f_c=100$ GHz, $B=10$ GHz, $|\theta_{\rm E}-\phi_{\rm B}|=0.01\pi$, $r_{\rm E}=0.2R_{\rm Ray}$ and $r_{\rm B}=1.1R_{\rm Ray}$. 
As shown
in Fig. \ref{Fig:sc_frequency}, the approximation in \eqref{Eq:NF_corre_approx} closely matches the
exact cross-correlation in \eqref{Eq:NF_corre_real}.
Moreover, the cross-correlation decreases monotonically with the subcarrier index, which is consistent with Remark \ref{Re:remark1}. Notably, the secrecy condition is violated for approximately half of the lower-frequency subcarriers, indicating that secure transmission cannot be achieved in this frequency range. In contrast, the remaining subcarriers satisfy the secrecy condition, thus enabling secure communication. Furthermore, the conventional narrowband secrecy threshold remains fixed and fails to ensure secure transmission under the same configuration. These observations align well with the theoretical insights presented in Remark \ref{Re:two_conditions}.}
\end{example}}
\begin{proposition}[New characteristics in the angle domain]\label{Le:new_angle}
    \emph{Given the location of the near-field eavesdropper $(\theta_{\rm E}, r_{\rm E})$, the cross-correlation $\rho(f_m,\theta_{\rm E}, r_{\rm E},\phi_{\rm B})$ is symmetric with respect to the legitimate user's angle $\phi_{\rm B}$ for any subcarrier $m$. Moreover, when the angles of the legitimate user and the eavesdropper are the same (i.e., $\theta_{\rm E}=\phi_{\rm B}$), the cross-correlation is higher at lower subcarrier frequencies and decreases with increasing frequency. The maximum cross-correlation, denoted by $\rho^{*}(f_1,\theta_{\rm E}, r_{\rm E},\phi_{\rm B})$, is attained at the lowest subcarrier frequency $f_1$.}
\end{proposition}
\begin{proof}
    The cross-correlation is symmetric with respect to $|\gamma_1|$, which depends on the angular difference, namely, $\cos{\theta_{\rm E}}-\cos{\phi_{\rm B}}$. Hence, it is symmetric with respect to this angular difference. When $\theta_{\rm E} = \phi_{\rm B}$, we have $\gamma_1 = 0$, and the only frequency-dependent term is $\gamma_2$, which increases with the subcarrier frequency $f_m$ through the factor $\frac{f_m}{f_c}$. Since the function $G(\cdot)$ generally decreases with increasing $\gamma_2$, the cross-correlation is maximized at the lowest subcarrier frequency $f_1$, and decreases as $f_m$ increases. This completes the proof. 
\end{proof}
\begin{proposition}[Secure transmission angle]\label{Le:sc_angle}
    \emph{Given the locations of a near-field eavesdropper $(\theta_{\rm E}, r_{\rm E})$ and a far-field legitimate user $(\phi_{\rm B}, r_{\rm B})$, the range of spatial angles over which secure transmission can be guaranteed on subcarrier $m$ is given by
    \begin{align}\label{Eq:secure_angle}
   \big\{\phi_{\rm B}\le \Xi_{\rm L}(&f_m,\theta_{\rm E}, r_{\rm E}, r_{\rm B})\big\}\nn\\
   &\cup\big\{\phi_{\rm B}\ge \Xi_{\rm R}(f_m,\theta_{\rm E}, r_{\rm E}, r_{\rm B})\big\},
    \end{align}
    where $\Xi_{\rm L}(f_m,\theta_{\rm E}, r_{\rm E}, r_{\rm B})$ and $\Xi_{\rm R}(f_m,\theta_{\rm E}, r_{\rm E}, r_{\rm B})$ are left-side and right-side solutions to $\rho(f_m,\theta_{\rm E}, r_{\rm E},\phi_{\rm B})=\frac{r_{\rm E}}{r_{\rm B}}$.
    }
\end{proposition}
\begin{proof}
    Given the threshold $\frac{r_{\rm E}}{r_{\rm B}}$, if the maximum cross-correlation $\rho^{*}(f_1,\theta_{\rm E}, r_{\rm E},\phi_{\rm B})>\frac{r_{\rm E}}{r_{\rm B}}$, $\rho(f_m,\theta_{\rm E}, r_{\rm E},\phi_{\rm B}) = \frac{r_{\rm E}}{r_{\rm B}}$ admits two solutions. These solutions, denoted as $\Xi_{\rm L}(f_m,\theta_{\rm E}, r_{\rm E}, r_{\rm B})$ and $\Xi_{\rm R}(f_m,\theta_{\rm E}, r_{\rm E}, r_{\rm B})$, represent angular boundaries beyond which the cross-correlation remains below the threshold, enabling secure transmission.  
    Thus, the secure angular region is the union of these two disjoint intervals as described in \eqref{Eq:secure_angle}, which completes the proof.
\end{proof}

\begin{proposition}[Secure transmission range]\label{Le:range}
    \emph{Given the location of a near-field eavesdropper $(\theta_{\rm E}, r_{\rm E})$ and the angle of a far-field legitimate user $\phi_{\rm B}$, the secure transmission range for the legitimate user on subcarrier $m$ is characterized by
        \begin{align}\label{Eq:secure_range}
    r_{\rm B}\le \Phi(f_m,\theta_{\rm E}, r_{\rm E},\phi_{\rm B}),
    \end{align}
    where $\Phi(f_m,\theta_{\rm E}, r_{\rm E},\phi_{\rm B})$ is the unique solution to the equation $\rho(f_m,\theta_{\rm E}, r_{\rm E},\phi_{\rm B})=\frac{r_{\rm E}}{r_{\rm B}}$.
    }
\end{proposition}
\begin{proof}
    The cross-correlation in \eqref{Eq:NF_corre_approx} is independent of the range of the far-field legitimate user. In contrast, the secrecy threshold $\frac{r_{\rm E}}{r_{\rm B}}$ decreases as the legitimate user's range $r_{\rm B}$ increases. Therefore, for fixed values of $(\theta_{\rm E}, r_{\rm E}, \phi_{\rm B})$, increasing $r_{\rm B}$ reduces the threshold and tightens the secrecy condition. The critical boundary at which the cross-correlation equals the threshold defines the maximum distance for secure transmission.
\end{proof}
  {\color{black}
\begin{example}
   \emph{In Figs. \ref{Fig:sc_angle} and \ref{Fig:sc_range}, we plot the secrecy condition versus the spatial angle and range of the legitimate user, respectively. 
   % The parameters are set to $f_c=100$ GHz, $B=10$ GHz, and $r_{\rm E}=0.2R_{\rm Ray}$. 
   % For the angular analysis, we use $\theta_{\rm E} = 0.5\pi$ and $r_{\rm B} = 1.1R_{\rm Ray}$, while for the range analysis, we set $|\theta_{\rm E} - \phi_{\rm B}| = 0.01\pi$. 
 From Fig. \ref{Fig:sc_angle}, 
   it is observed that the approximation in \eqref{Eq:NF_corre_approx} closely agrees with the exact cross-correlation in \eqref{Eq:NF_corre_real} over all examined subcarriers.
   We observe that the cross-correlation on three representative subcarriers, namely, the first subcarrier $f_1$, the central subcarrier $f_c$, and the last subcarrier $f_M$, is symmetric with respect to the legitimate user's angle, and reaches its maximum when $\theta_{\rm E}=\phi_{\rm B}$. Additionally, among these subcarriers, the lowest-frequency subcarrier yields the highest cross-correlation, while the highest-frequency subcarrier yields the lowest one. These observations are consistent with the results presented in Proposition \ref{Le:new_angle}. More importantly, the angles over which the secrecy condition is satisfied occur in two separate angular regions, which aligns with the analytical development in Proposition \ref{Le:sc_angle}. Fig.~\ref{Fig:sc_range} further explores the impact of the legitimate user’s range on secure transmission. It is observed that the conditions for secure transmission become more stringent as the distance of the legitimate user increases, due to the decreasing secrecy threshold $\frac{r_{\rm E}}{r_{\rm B}}$. In contrast, the cross-correlation values for the three representative subcarriers remain constant, as they are independent of the range of the legitimate user. Notably, the secure transmission range varies across subcarriers: the lower-frequency subcarrier exhibits the narrower secure range, while the higher-frequency subcarrier allows for the broader ones. This difference arises from the fact that the cross-correlation is higher at lower frequencies and lower at higher frequencies. These observations are consistent with the results in Remark \ref{Re:remark1} and Proposition \ref{Le:range}.}
\end{example}}
{\color{black}
\begin{example}[Insecure transmission regions: mixed-field versus far-field wideband systems]
    \emph{In Fig. \ref{Fig:sc_2dregion}, we plot the insecure transmission regions of the central subcarrier $f_c$ and the last subcarrier $f_M$ for the far-field legitimate user under the setting  $\theta_{\rm E}=0.5\pi$ and $r_{\rm E}=0.2R_{\rm Ray}$. First, it can be observed that the insecure region shrinks with the subcarrier index, which is consistent with Proposition \ref{Le:secure_sub}. Second, the insecure regions across subcarriers exhibit clear symmetry, in line with the results of Proposition \ref{Le:sc_angle}. Next, the insecure region in the range domain becomes larger as the legitimate user’s range increases, which agrees well with Proposition \ref{Le:range}. Finally, the insecure regions in mixed-field wideband systems are broader than those in far-field systems due to the inherent mixed-field energy spread effect.}
\end{example}}
\begin{remark}
    \emph{To summarize, it is important to highlight that, despite the preceding analysis of several key factors, namely, the subcarrier index, legitimate user's spatial angle, legitimate user's range, the secrecy condition remains difficult to satisfy under various system configurations. Specifically, as illustrated in Figs. \ref{Fig:sc_frequency}--\ref{Fig:sc_range}, secure transmission is achieved over approximately half of the subcarriers; the secure angular regions are confined to  $[-1,-0.22]$ rad and $[0.22,1]$ rad; and the secure transmission range on the lowest-frequency subcarrier $f_1$ is restricted to $[R_{\rm Ray},1.05R_{\rm Ray}]$, all of which indicate relatively narrow secure regions. These limitations underscore the necessity of incorporating AN into mixed-field wideband systems to enhance the PLS. In the following subsection, we further investigate the role and effectiveness of \emph{frequency-selective} AN on secure transmission performance in mixed-field wideband scenarios.}
\end{remark}
\subsection{Mixed-Field Wideband Secure Transmission with AN}
We now consider secure transmission in mixed-field wideband systems with the incorporation of AN. The associated secrecy condition, which ensures secure communication in the presence of AN, is presented in the following lemma. For notational simplicity, we denote $\rho(f_m)$ as a shorthand for $\rho(f_m,\theta_{\rm E}, r_{\rm E},\phi_{\rm B})$ in the remainder of this section.
{\color{black}
\begin{lemma}\label{Le:scwithAN}
    \emph{For any subcarrier $m\in\mathcal{M}$ where the original secrecy condition in \eqref{Eq:Wideband} is not met (i.e., secure transmission cannot be guaranteed), expressed as 
    \begin{equation}\label{Eq:sc_withoutAN}
       r^2_{\rm E}- r^2_{\rm B} \rho^2(f_m)< 0,
    \end{equation}
    secure transmission can be achieved by incorporating \emph{frequency-selective} AN, provided that the following condition holds:
    \begin{equation}\label{Eq:sc_withAN}
        r^2_{\rm E}- r^2_{\rm B} \rho^2(f_m)> -\frac{PNc^2(1-\rho^4(f_m))}{16f_m^2\pi^2\sigma^2}.
    \end{equation}
    Under this condition, secure transmission over subcarrier $m$ can be reestablished by allocating a minimum power $P_{\rm E, min}$ to AN, given by
    \begin{equation}\label{Eq:minimum_ANpower}
        P_{\rm E, min} = \frac{16f_m^2\pi^2\sigma^2(r^2_{\rm B} \rho^2(f_m)-r^2_{\rm E})}{Nc^2(1-\rho^4(f_m))} <P.
    \end{equation}
    }
\end{lemma}
\begin{proof}
    Please refer to Appendix \ref{App1}.
\end{proof}
\begin{remark}[The effect of frequency-selective AN in mixed-field wideband PLS]
   \emph{Lemma \ref{Le:scwithAN} offers the following insights:
   \begin{itemize}
       \item The incorporation of frequency-selective AN allows the transformation of the previously insecure subcarriers into secure ones. Specifically, the secrecy condition becomes less stringent:  the threshold is relaxed from $0$ (as in \eqref{Eq:sc_withoutAN}) to a negative bound (as in \eqref{Eq:sc_withAN}). This threshold is significantly influenced by factors such as the number of antennas, the per-subcarrier transmit power, the subcarrier frequency, and the locations of both the legitimate user and the eavesdropper.
       \item The required AN power is determined by the frequency band and is notably independent of the per-subcarrier transmit power. 
       \item Moreover, Lemma~\ref{Le:scwithAN} implies that the secrecy rate on each subcarrier is upper-bounded, as it first increases and then decreases with AN power. This upper bound can be closely approached through well-designed power allocation strategies, as shown in the next lemma.
   \end{itemize}
   } 
\end{remark}

  \begin{figure*}[t!]
\begin{minipage}{.3\textwidth}
	\centering
\includegraphics[width=1\columnwidth]{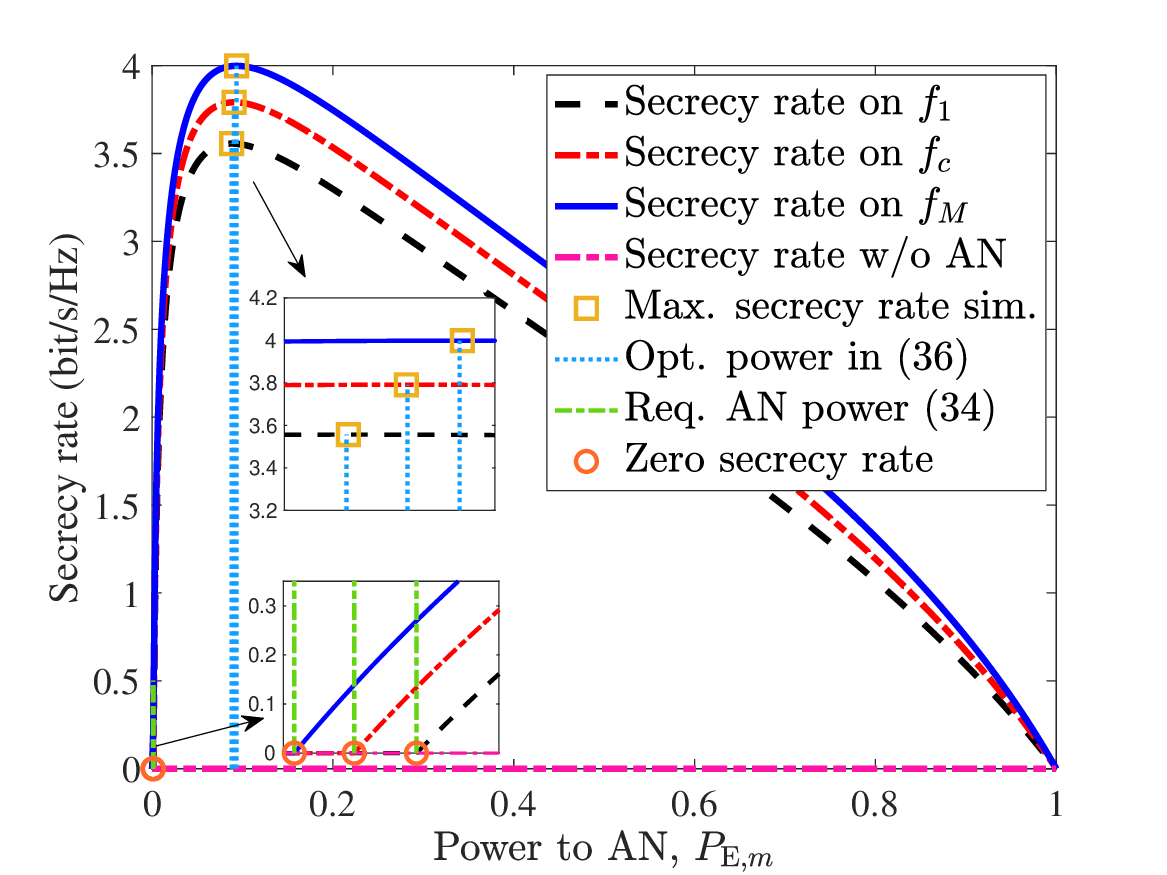}
	\caption{{\textbf{Case 1:} Data transmission is insecure without AN (i.e., \eqref{Eq:Wideband} does not hold) but becomes secure with AN (i.e., \eqref{Eq:sc_withAN} holds).}\label{Fig:case1}} 
    \end{minipage}	
    \hfill
    \begin{minipage}{.3\textwidth}
	\centering
\includegraphics[width=1\columnwidth]{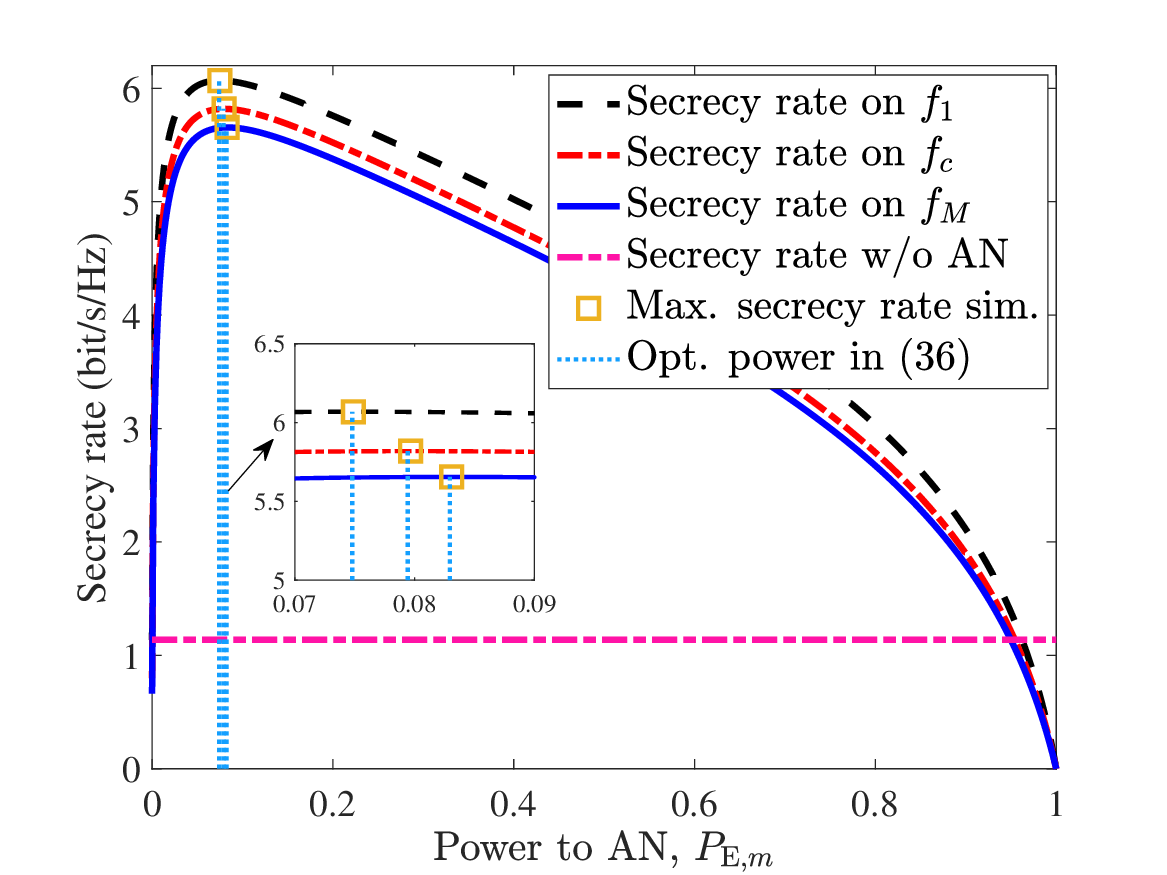}
	\caption{{\textbf{Case 2:} Data transmission is secure without AN (i.e., \eqref{Eq:Wideband} holds) and AN further enhances the performance (i.e., \eqref{Eq:sec_max} holds).}\label{Fig:case2}} 
    \end{minipage}	
    \hfill
        \begin{minipage}{.3\textwidth}
\centering
\includegraphics[width=1\columnwidth]{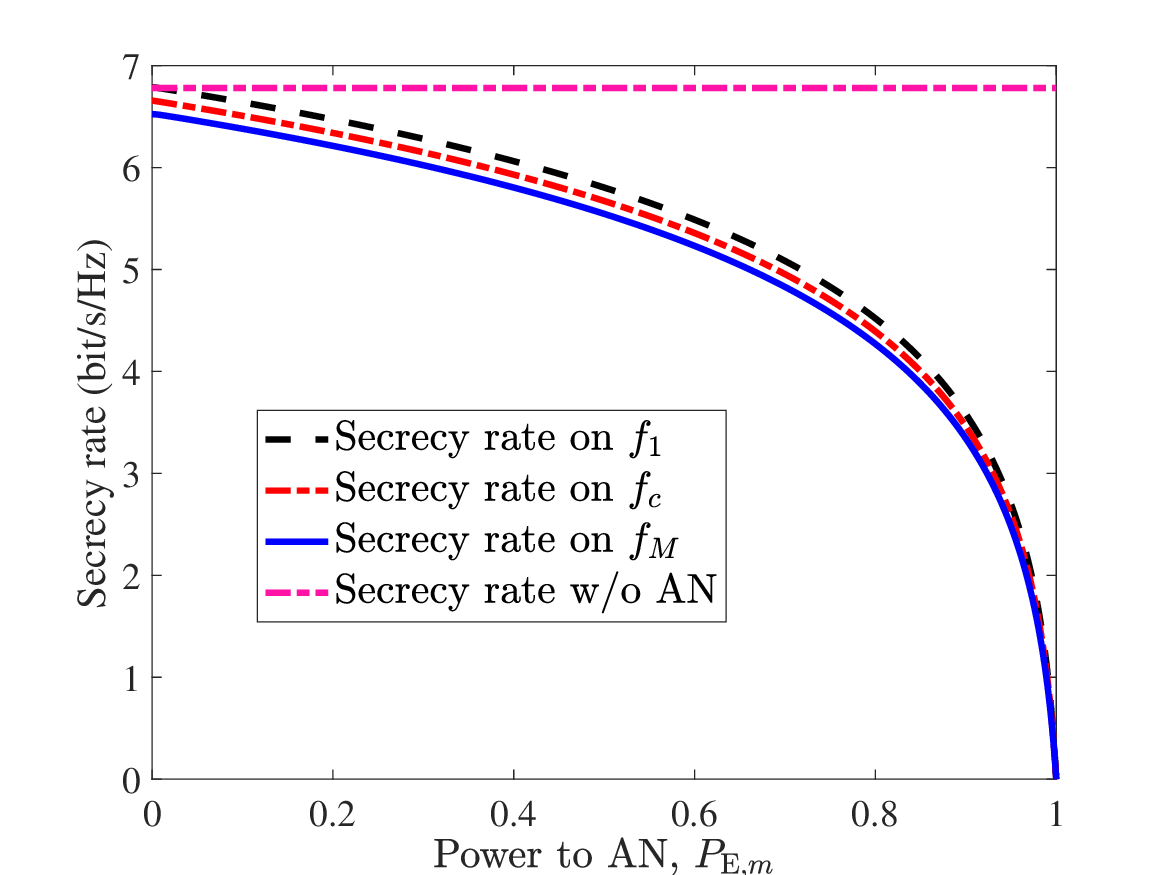}
	\caption{{\textbf{Case 3:} Data transmission with AN is not beneficial as the condition in \eqref{Eq:sec_max} is not satisfied.}\label{Fig:case3}} 
    \end{minipage}
    \vspace{-18pt}
\end{figure*}
\begin{lemma}\label{Le:opt_pow}
	\emph{The optimal power allocation that maximizes the secrecy rate $R^{\rm Sec}_{{\rm B},m}$ over subcarrier $m$ in \eqref{Eq:SC_rate_expression1} is given as:
		\begin{itemize}
  			\item If the following condition holds:
               \begin{align}\label{Eq:sec_max}
       r^2_{\rm E}- r^2_{\rm B} \rho^2(f_m)< &\frac{P^2 N^2c^4\rho^2(f_m)( r^2_{{\rm B}}-r^2_{{\rm E}}\rho^2(f_m))}{(16r_{{\rm B}}r_{{\rm E}}f^2_m\pi^2\sigma^2)^2}\nn\\
       &+\frac{PNc^2\rho^2(f_m)(r^4_{{\rm B}}-r^4_{{\rm E}})}{16f^2_m\pi^2\sigma^2},
   \end{align}
            then the secrecy rate first increases  with $P_{{\rm E},m}$ and then decreases. The maximum is achieved at:
			\begin{equation}\label{Eq:pow_max}
				\begin{cases}
					P_{{\rm E},m}= -\frac{\Upsilon_2}{{2\Upsilon_1}}-\frac{\sqrt{{\Upsilon_2^2}-4{\Upsilon_1}\Upsilon_3}}{2\Upsilon_1},\\
					P_{{\rm B},m}=P-P_{{\rm E},m},
				\end{cases}
			\end{equation}
   	where the coefficients $\Upsilon_1$, $\Upsilon_2$, and $\Upsilon_3$ are given in \eqref{Eq:coe} at the bottom of the next page.
    \begin{figure*}[b]
    \vspace{-16pt}
    \hrulefill
            \begin{align}\label{Eq:coe}
\Upsilon_1 &= \dfrac{(N c^2)^3}{(16 f_m^2 \pi^2)^4 r_{\rm B}^4 r_{\rm E}^4} \left[ P (N c^2) \rho^2(f_m) (\rho^4(f_m) - 1) + (16 f_m^2 \pi^2) (\rho^2(f_m) - 1) \sigma^2 (r_{\rm B}^2 + \rho^4(f_m) r_{\rm E}^2) \right], \nn\\
\Upsilon_2 &= \dfrac{2 (N c^2)^2}{(16 f_m^2 \pi^2)^3 r_{\rm B}^4 r_{\rm E}^4} \left[ P (N c^2) (r_{\rm B}^2 + r_{\rm E}^2) (\rho^2(f_m) - 1) \rho^2(f_m) \sigma^2 + (16 f_m^2 \pi^2) (\rho^4(f_m) - 1) \sigma^4 r_{\rm B}^2 r_{\rm E}^2 \right], \nn\\
\Upsilon_3 &= \dfrac{(N c^2)}{(16 f_m^2 \pi^2)^3 r_{\rm B}^4 r_{\rm E}^4} \big[ P^2 (N c^2)^2 (r_{\rm B}^2 - \rho^2(f_m) r_{\rm E}^2) \rho^2(f_m) \sigma^2 + P (N c^2) (16 f_m^2 \pi^2) (r_{\rm B}^4 - r_{\rm E}^4) \rho^2(f_m) \sigma^4\nn\\
&\quad\quad\quad\quad\quad\quad\quad\quad+ (16 f_m^2 \pi^2)^2 (r_{\rm B}^2 \rho^2(f_m) - r_{\rm E}^2) \sigma^6 \big].
\end{align}
    \end{figure*}
			\item Otherwise,
          the secrecy rate monotonically decreases with $P_{{\rm E},m}$, and the optimal solution is:
		\begin{equation}\label{Eq:pow_mon}
			\begin{cases}
				P_{{\rm E},m}= 0,\\
				P_{{\rm B},m}=P.
			\end{cases}
		\end{equation}
		\end{itemize}}
\end{lemma}
 
\begin{proof}
   Please refer to Appendix \ref{app2}.
\end{proof}

Lemma \ref{Le:opt_pow} reveals a noteworthy insight into the role of \emph{frequency-selective} AN in enhancing mixed-field wideband PLS. Specifically, in the absence of AN, it follows directly from Lemma \ref{Le:Narrowband} that the secrecy rate monotonically increases with the transmit power allocated to the legitimate user, provided that the secrecy condition in \eqref{Eq:NarrowbandSC} is satisfied. In contrast, when frequency-selective AN is introduced, it is worth highlighting that the threshold in \eqref{Eq:sec_max} (i.e., the right side of \eqref{Eq:sec_max}), which determines the point at which the secrecy rate changes from increasing to decreasing, is strictly positive. This suggests an interesting result that even when the wideband secrecy condition in \eqref{Eq:Wideband} is already satisfied, a non-zero AN power may still further improve the secrecy rate, as the optimal power allocation may lie within a regime where AN remains beneficial. In other words, AN not only facilitates secure transmission on subcarriers where the secrecy condition fails, but can also enhance performance on subcarriers that are already secure. This result underscores the broader utility of frequency-selective AN in mixed-field wideband PLS systems and aligns with similar observations demonstrated in narrowband systems (e.g., \cite{yunpusecure}).}

\begin{example}\label{Ex:AN}
    \emph{Lemmas \ref{Le:scwithAN} and \ref{Le:opt_pow} analytically demonstrate that the {frequency-selective} AN not only secures mixed-field PLS but also enhances its secrecy performance. To validate these findings, we present a concrete numerical example. 
    % Specifically, we consider a mixed-field wideband PLS system with $N=256$, $f_c=100$ GHz, $B=10$ GHz, and $M=11$. 
    Figs. \ref{Fig:case1}--\ref{Fig:case3} illustrate the secrecy rates of three representative subcarriers, i.e., the first $f_1$, the central $f_c$, and the last $f_M$, versus the AN power. The observations can be categorized into three cases:
    \begin{itemize}
        \item \textbf{Case 1:} Here, the secrecy condition in \eqref{Eq:Wideband} is not met, but \eqref{Eq:sc_withAN} holds. As shown in Fig. \ref{Fig:case1}, the secrecy rates of all three subcarriers first increase with $P_{{\rm E},m}$, reach a peak and then decrease. Even a small amount of AN power turns previously insecure subcarriers secure, aligning with the theoretical minimum in \eqref{Eq:minimum_ANpower}. The observed optimal AN power closely matches the analytical solution in \eqref{Eq:pow_max}, with maximum secrecy rates marked by squares. Notably, $f_M$ achieves the highest secrecy rate due to its highest frequency, which results in the weakest cross-correlation (i.e., minimal information leakage), and thus the greatest secrecy improvement. This is consistent with Remark \ref{Re:two_conditions}.
        \item \textbf{Case 2:} In this case, both \eqref{Eq:Wideband} and \eqref{Eq:sec_max} are satisfied. While secure transmission is ensured, Fig. \ref{Fig:case2} shows that moderate AN power further improves secrecy rates, following the same trend as in  \textbf{Case 1}. This highlights that allocating power to AN generally enhances secrecy in mixed-field wideband systems, as frequency-selective AN exploits the beam-focusing effect to degrade the eavesdropper’s reception. Interestingly, $f_1$ now achieves the highest secrecy rate, which is contrary to \textbf{Case 1}. With secure transmission ensured and information leakage limited, secrecy performance is mainly determined by path loss, favoring the lower-frequency subcarriers that experience reduced path loss.
        \item \textbf{Case 3:} Here, the condition in \eqref{Eq:sec_max} is not satisfied. It is clearly observed from Fig. \ref{Fig:case3} that the optimal strategy across all subcarriers is to allocate no power to AN, consistent with Lemma \ref{Le:opt_pow}. Since information leakage is already negligible, dedicating all power to information transmission yields the best secrecy performance.
    \end{itemize}
    }
\end{example}
\begin{remark}[How to design the AN beamformer?]\label{Re:ANdesign}
    \emph{It is worth highlighting that the amount of power allocated to AN, whether for securing the transmission or enhancing the performance of mixed-field wideband PLS, is typically marginal, as illustrated in Example \ref{Ex:AN}. This is largely attributed to the highly effective near-field beam-focusing capability, which allows AN to substantially impair eavesdropping \cite{wang2023beamfocusing}. Moreover, it has been shown in \cite{yunpusecure} that the interference caused by AN to far-field legitimate users is nearly negligible, owing to the significantly higher path loss in the far field. Based on these insights, we conclude that the AN beamformer should be carefully directed toward the near-field eavesdropper to achieve near-optimal secrecy performance under the optimal power allocation. These observations motivate the low-complexity design introduced in Section \ref{Sec:low}, which will be numerically validated in Section \ref{Sec:NumericalR}.}
\end{remark}
\vspace{-12pt}
\section{Proposed Solution to Problem (P1)}\label{Sec:general}
{\color{black}In this section, we first propose a two-stage hybrid beamforming framework to obtain an effective suboptimal solution to problem (P1). Then, we develop a low-complexity algorithm based on the analytical insights obtained in Section \ref{Sec:Special_case}.}
\vspace{-10pt}
\subsection{Proposed Two-stage Hybrid Beamforming Approach}\label{Sec:conventional}
\subsubsection{Composite Analog beamformer} We begin with the design of the composite analog beamformer $\mathbf{F}_{{\rm A},m}$, which employs both TTD and PS components. For each legitimate user $k$ on subcarrier $m$, the $k$-th column of the analog beamformer $\mathbf{F}_{{\rm A},m}$ is designed to maximize its received signal power. In contrast, for the eavesdropper, the associated analog beamformer is determined via maximizing the received AN power. Accordingly, based on the individual LoS path of the far-field legitimate users and the near-field eavesdropper, the composite analog beamformer ${\mathbf{F}}_{{\rm A},m}$ is constructed as:
\begin{equation}
    {\mathbf{F}}_{{\rm A},m} = \sqrt{N}[\mathbf{a}(f_m,\phi_{{\rm B},1}),\ldots,\mathbf{a}(f_m,\phi_{{\rm B},K}),\mathbf{b}(f_m,\theta_{\rm E}, r_{\rm E})],
\end{equation}
where
$\mathbf{F}^{\rm (PS)}_{{\rm A}} = \sqrt{N}[\mathbf{a}(f_c,\phi_{{\rm B},1}),\ldots,\mathbf{a}(f_c,\phi_{{\rm B},K}),\mathbf{b}(f_c,\theta_{\rm E}, r_{\rm E})]$ and $\mathbf{F}^{\rm (TTD)}_{{\rm A},m} = [e^{-j{2\pi f_m}\mathbf{t}_{{\rm B},1}},\ldots,e^{-j{2\pi f_m}\mathbf{t}_{{\rm B},K}},e^{-j{2\pi f_m}\mathbf{t}_{\rm E}}]$, with $\mathbf{t}_{{\rm B},k} = \frac{\left[\angle{\mathbf{a}(f_c,\phi_{{\rm B},k})}-\angle{\mathbf{a}(f_m,\phi_{{\rm B},k})})\right]}{2\pi f_m}$ and $\mathbf{t}_{\rm E} = \frac{\left[\angle{\mathbf{b}(f_c,\theta_{\rm E}, r_{\rm E})}-\angle{\mathbf{b}(f_m,\theta_{\rm E}, r_{\rm E})})\right]}{2\pi f_m}$. 

\subsubsection{Digital beamformer}\label{sub:dig} Given the designed analog beamformer ${\mathbf{F}}_{{\rm A},m}$, we then optimize the digital beamformers ${\mathbf{F}}_{{\rm D},m}$ and ${\mathbf{f}}_{{\rm N},m}$ on subcarrier $m$, based on the effective channels for all legitimate users and the eavesdropper, which are given by
\begin{align}
     \Bar{\mathbf{h}}_{{\rm B},m,k} = {\mathbf{F}}^{H}_{{\rm A},m}{\mathbf{h}}_{{\rm B},m,k},~~ 
     \Bar{\mathbf{h}}_{{\rm E},m}  = {\mathbf{F}}^{H}_{{\rm A},m}{\mathbf{h}}_{{\rm E},m}.
\end{align}
The achievable secrecy rate of the legitimate user $k$ on subcarrier $m$ is then expressed as
\begin{align}
    & R^{\rm Sec}_{{\rm B},m,k} = \nn\\
    &\!\!\log_{2}\!\left(\! 1\!+\!\frac{| \Bar{\mathbf{h}}^H_{{\rm B},m,k}\mathbf{f}_{{\rm D},m,k}| ^2}{\sum^{K}_{i=1,i\neq k}|  \Bar{\mathbf{h}}^H_{{\rm B},m,k}\mathbf{f}_{{\rm D},m,i}| ^2+| \Bar{\mathbf{h}}_{{\rm B},m,k}^H\mathbf{f}_{{\rm N},m}|^2+\sigma^2_{{\rm B},m,k}}\!\right)\nn\\
    &-\log_{2} \left(1+\frac{|\Bar{\mathbf{h}}^H_{{\rm E},m}\mathbf{f}_{{\rm D},m,k}|^2}{|\Bar{\mathbf{h}}_{{\rm E},m}^H\mathbf{f}_{{\rm N},m}|^2+\sigma^2_{{\rm E},m}} \right).
\end{align}
Thus, problem (P1) can be equivalently formulated as
\begin{subequations}
	\begin{align}
		({\bf P2}):\max_{\substack{\{\mathbf{F}_{{\rm D},m},\mathbf{f}_{{\rm N},m}\} }}  \frac{1}{M}\sum^{M}_{m=1}\sum^{K}_{k=1}R^{\rm Sec}_{{\rm B},m,k},~~
		\text{s.t.}~
		\eqref{P1:pow_cons}.\nn
	\end{align}
 \end{subequations}
Problem (P2) remains non-convex due to the coupling between $\mathbf{F}_{{\rm D},m}$ and $\mathbf{f}_{{\rm N},m}$ in its objective function. To address this challenge, we develop an efficient algorithm to obtain a high-quality solution to problem (P2) by leveraging the SDR and SCA. Specifically, let $\mathbf{H}_{{\rm B},m,k} = \Bar{\mathbf{h}}_{{\rm B},m,k}\Bar{\mathbf{h}}_{{\rm B},m,k}^H$, $\mathbf{H}_{{\rm E},m} = \Bar{\mathbf{h}}_{{\rm E},m}\Bar{\mathbf{h}}_{{\rm E},m}^H$,  $\mathbf{W}_{{\rm D},m,k} = {\mathbf{f}}_{{\rm D},m,k}{\mathbf{f}}_{{\rm D},m,k}^H$, and $\mathbf{W}_{{\rm N},m} = {\mathbf{f}}_{{\rm N},m}{\mathbf{f}}_{{\rm N},m}^H$, problem (P2) can be equivalently reformulated as
\begin{subequations}
	\begin{align}
		({\bf P3}):\min_{\substack{\{\mathbf{W}_{{\rm D},m,k},\\\mathbf{W}_{{\rm N},m}\}}}  &\frac{1}{M}\sum^{M}_{m=1}\left(N_{m,1}+N_{m,2}-D_{m,1}-D_{m,2}\right)\nn
		\\
		\text{s.t.}~~
		&	\eqref{P1:pow_cons},\nn\\
            & \mathbf{W}_{{\rm D},m,k} \succeq 0, \forall{m} \in \mathcal{M}, k \in \mathcal{K},\label{P3:sdr}\\
            & \mathbf{W}_{{\rm N},m} \succeq 0, \forall{m} \in \mathcal{M},\\
            & \operatorname{Rank}(\mathbf{W}_{{\rm D},m,k})\le1,\forall{m} \in \mathcal{M}, k \in \mathcal{K},\label{P3:rank}\\
            &\operatorname{Rank}(\mathbf{W}_{{\rm N},m})\le1,\forall{m} \in \mathcal{M},\label{P3:rank1}
	\end{align}
 \end{subequations}
where the terms $N_{m,1}$, $N_{m,2}$, $D_{m,1}$, and $D_{m,2}$ are defined as
\begin{align}\label{Eq:P3_Coe}
        N_{m,1}  &=-\sum^{K}_{k=1}\log_2\bigg(\sum^{K}_{i=1}\mathrm{Tr}(\mathbf{H}_{{\rm B},m,k}\mathbf{W}_{{\rm D},m,k})\nn\\
        &\quad\quad\quad\quad\quad\quad+\mathrm{Tr}(\mathbf{H}_{{\rm B},m,k}\mathbf{W}_{{\rm N},m})+\sigma^2_{{\rm B},m,k}\bigg), \\
        N_{m,2} &= -K\log_2\left(\mathrm{Tr}(\mathbf{H}_{{\rm E},m}\mathbf{W}_{{\rm N},m})+\sigma^2_{{\rm E},m}\right),\\
        D_{m,1} & = -\sum^K_{k=1}\log_2\bigg(\sum^{K}_{i=1,i\neq k}\mathrm{Tr}(\mathbf{H}_{{\rm B},m,k}\mathbf{W}_{{\rm D},m,i})\nn\\
         &\quad\quad\quad\quad\quad\quad+\mathrm{Tr} (\mathbf{H}_{{\rm B},m,k}\mathbf{W}_{{\rm N},m})+\sigma^2_{{\rm B},m,k}\bigg),\\
        D_{m,2} &= -\sum^K_{k=1}\log_2\bigg(\mathrm{Tr}(\mathbf{H}_{{\rm E},m}\mathbf{W}_{{\rm D},m,k})\nn\\
        &\quad\quad\quad\quad\quad\quad+\mathrm{Tr}(\mathbf{H}_{{\rm E},m}\mathbf{W}_{{\rm N},m})+\sigma^2_{{\rm E},m}\bigg).
\end{align}
We first construct a convex upper bound for the objective function in an iterative manner by deriving global underestimators for the functions $D_{m,1}$ and $D_{m,2}$ based on their first-order Taylor approximations. For any feasible point $\mathbf{W}_{{\rm D},m}^{(t)} =\{\mathbf{W}^{(t)}_{{\rm D},m,k}\}^{K}_{k=1}$ and $\mathbf{W}^{(t)}_{{\rm N},m}$, the lower bound of $D_{m,1}$ is given by
\begin{align}
    &D_{m,1}\left(\mathbf{W}_{{\rm D},m},\mathbf{W}_{{\rm N},m}\right)\ge D_{m,1}\left(\mathbf{W}^{(t)}_{{\rm D},m},\mathbf{W}^{(t)}_{{\rm N},m}\right)\nn\\
    &\!\!+\sum^{K}_{k=1}\!\mathrm{Tr}\!\left(\nabla^H_{\mathbf{W}_{{\rm D},m,k}}\!D_{m,1}\!\left(\!\mathbf{W}^{(t)}_{{\rm D},m},\!\mathbf{W}^{(t)}_{{\rm N},m}\!\right)\!\left(\!\mathbf{W}_{{\rm D},m,k}\!-\!\mathbf{W}^{(t)}_{{\rm D},m,k}\!\right)\right)\nn\\
    &+\mathrm{Tr}\left(\nabla^H_{\mathbf{W}_{{\rm N},m}}D_{m,1}\left(\mathbf{W}^{(t)}_{{\rm D},m},\mathbf{W}^{(t)}_{{\rm N},m}\right)\left(\mathbf{W}_{{\rm N},m}-\mathbf{W}^{(t)}_{{\rm N},m}\right)\right)\nn\\
    &\triangleq \tilde{D}_{m,1}\left(\mathbf{W}_{{\rm D},m},\mathbf{W}_{{\rm N},m}\right),
\end{align}
where the gradients of $D_{m,1}$ with respect to $\mathbf{W}_{{\rm D},m,k}$ and $\mathbf{W}_{{\rm N},m}$ are given in \eqref{Eq:gradient} at the top of the next page. The lower bound of $D_{m,2}$ and its corresponding gradients can be derived in a similar manner and are omitted for brevity.
\begin{figure*}[t]
    \begin{align}\label{Eq:gradient}
        \nabla_{\mathbf{W}_{{\rm D},m,k}}D_{m,1}(\mathbf{W}_{{\rm D},m,k},\mathbf{W}_{{\rm N},m})&=-\frac{1}{\ln2}\sum^{K}_{j=1,j\neq k}\frac{\mathbf{H}_{{\rm B},m,j}}{\sum^{K}_{i=1,i\neq j}\mathrm{Tr}(\mathbf{H}_{{\rm B},m,j}\mathbf{W}_{{\rm D},m,i})+\mathrm{Tr} (\mathbf{H}_{{\rm B},m,j}\mathbf{W}_{{\rm N},m})+\sigma^2_{{\rm B},m,j}},\nn\\
        \vspace{-5pt}
        \nabla_{\mathbf{W}_{{\rm N},m}}D_{m,1}(\mathbf{W}_{{\rm D},m,k},\mathbf{W}_{{\rm N},m})&=-\frac{1}{\ln2}\sum^{K}_{j=1}\frac{\mathbf{H}_{{\rm B},m,j}}{\sum^{K}_{i=1,i\neq j}\mathrm{Tr}(\mathbf{H}_{{\rm B},m,j}\mathbf{W}_{{\rm D},m,i})+\mathrm{Tr} (\mathbf{H}_{{\rm B},m,j}\mathbf{W}_{{\rm N},m})+\sigma^2_{{\rm B},m,j}}.
    \end{align}
     \vspace{-5pt}
    \hrulefill
     \vspace{-12pt}
\end{figure*}
Therefore, for any given $\mathbf{W}_{{\rm D},m}^{(t)} =\{\mathbf{W}^{(t)}_{{\rm D},m,k}\}^{K}_{k=1}$ and $\mathbf{W}^{(t)}_{{\rm N},m}$, an upper bound of the objective function in problem (P3) can be obtained by solving the following optimization problem:
\begin{subequations}
	\begin{align}
		({\bf P4}):\min_{\substack{\{\mathbf{W}_{{\rm D},m,k},\\\mathbf{W}_{{\rm N},m}\}}}  &\frac{1}{M}\sum^{M}_{m=1}\left(N_{m,1}+N_{m,2}-\tilde{D}_{m,1}-\tilde{D}_{m,2}\right)\label{P4:obj}\nn
		\\
		\text{s.t.}~~~~
		&	\eqref{P1:pow_cons},\eqref{P3:sdr}-\eqref{P3:rank1},\nn
	\end{align}
 \end{subequations}
The only remaining non-convexity in problem (P4) arises from the rank constraints in \eqref{P3:rank} and \eqref{P3:rank1}. To address this, the SDR technique is employed to eliminate the rank constraints. The resulting problem becomes convex with respect to all optimization variables and can thus be efficiently solved using standard convex program solvers such as CVX. Moreover, the tightness of SDR has been rigorously established in the literature and is therefore omitted here for brevity \cite{8723525}.

{
\begin{remark}[Algorithm convergence and computational complexity]
    \emph{First, it is worth noting that the minimum values of problem (P4) serve as the upper bound for the optimal values of problem (P3). By performing the developed algorithm, we can monotonically tighten the upper bound, such that the objective function of problem (P4) is non-increasing in each iteration. This ensures that the proposed algorithm converges to a stationary point of problem (P3) \cite{10005137}. Furthermore, its computational complexity is given by $\mathcal{O}(\sqrt{(K+1)MK}\left((K+1)M^3K^2+(K+1)M^2K^3\right)\log\frac{1}{\epsilon})$, where $\epsilon$ is a predefined accuracy \cite{6891348}.} 
\end{remark}}

% {\color{black}
% \begin{remark}[On low-complexity beam synthesis for wideband XL-array systems]
%     \emph{While the beamforming strategies derived in this paper provide theoretical performance benchmarks for
% mixed-field wideband PLS, we note that full per-subcarrier
% optimization may be computationally demanding for XL-arrays. A practical and scalable alternative is to employ a codebook-based beam synthesis framework, inspired by the advanced Type-II architecture in modern cellular standards \cite{9762861}. By augmenting the standard spatial basis to include both far-field angular vectors and precomputed near-field focusing vectors, secure waveforms can be constructed via efficient, closed-form linear combinations rather than iterative optimization. Specifically: (i) information beams are synthesized by linearly combining user-aligned and eavesdropper-aligned bases to create deep suppression zones at the eavesdropper; (ii) AN is generated by focusing energy on the eavesdropper while orthogonalizing the beam against the user’s channel. Crucially, for wideband scalability, the frequency-selectivity of these combining weights is compressed using a frequency-domain basis \cite{9762861}, which decouples the processing overhead from the number of subcarriers.}
% \end{remark}
% }
   \vspace{-15pt}
\subsection{Proposed Low-complexity Hybrid Beamforming Approach}\label{Sec:low}
It should be highlighted that the computational complexity of the previously developed two-stage algorithm is significantly influenced by both the number of users and the number of subcarriers, which becomes computationally expensive when either parameter size is large. To address this challenge, we propose an efficient \emph{low-complexity} approach in this subsection, which is developed based on the obtained analytical insights in Section \ref{Sec:Special_case}. 

{\color{black}
Specifically, the proposed \emph{low-complexity} approach retains the two-stage hybrid beamforming architecture but introduces a simplified design for the AN beamformer. As demonstrated in Remark~\ref{Re:ANdesign}, designing the AN beamformer on each subcarrier in an MRT manner towards the near-field eavesdropper can effectively degrade its eavesdropping while causing minimal interference to far-field legitimate users \cite{yunpusecure}. This strategy results in negligible performance loss compared to the AN design presented in Section \ref{sub:dig}. Accordingly, for the design of AN on each subcarrier, the corresponding hybrid beamformer is constructed to steer a beam that precisely aligns with the LoS channel of the near-field eavesdropper on each subcarrier $m$, formulated as
% \begin{equation}
%     \mathbf{F}_{{\rm A},m}\mathbf{f}_{{\rm N},m}=\sqrt{P_{{\rm E},m}}\frac{\mathbf{h}_{{\rm E}, m}}{\|\mathbf{h}_{{\rm E}, m}\|},
% \end{equation}
\begin{equation}
   \mathbf{F}_{{\rm A},m}\mathbf{f}_{{\rm N},m}=\sqrt{P_{{\rm E},m}}\mathbf{b}(f_m,\theta_{\rm E}, r_{\rm E}),
\end{equation}
where $P_{{\rm E},m}$ denotes the AN power allocated to subcarrier $m$, which is determined using the same method as in the original two-stage hybrid beamforming approach.}

{
\begin{remark}
    \emph{The computational complexity of the proposed low-complexity approach is now reduced to $\mathcal{O}\left(\sqrt{MK^2}\left(M^3K^3+M^2K^4\right)\log\frac{1}{\epsilon}\right)$. For instance, when the number of subcarriers is $M=513$ and the number of legitimate users is $K=3$, this approach reduces the overall complexity by approximately $35\%$ compared with the proposed two-stage algorithm in Section \ref{Sec:conventional}.}
    % The performance comparison between these two approaches will be illustrated in Section \ref{Sec:NumericalR} }
\end{remark}}

   \vspace{-16pt}
\section{Numerical Results}\label{Sec:NumericalR}
In this section, numerical results are provided to demonstrate the new characteristic of the investigated mixed-field wideband PLS systems, as well as the effectiveness of the two proposed hybrid beamforming algorithms.

   \vspace{-10pt}
\subsection{System Setting and Benchmark Schemes}
The system parameters are
configured as follows. The BS is equipped with $N=256$ antennas with half-wavelength spacing and employs $N_{\rm RF}=4$ RF chains. The central frequency is set to $f_c = 100$ GHz, with an overall system bandwidth of $B=10$ GHz and $M=513$ subcarriers \cite{10004962}. There are $K=3$ far-field legitimate users, with angles randomly sampled from the range $[\frac{\pi}{3},\frac{2\pi}{3}]$, and their distances are randomly selected from $R_{\rm Ray}$ to $1.5R_{\rm Ray}$. The near-field eavesdropper is positioned within the angular range $[\frac{\pi}{3},\frac{2\pi}{3}]$ and distance range $[0.1R_{\rm Ray},0.5R_{\rm Ray}]$. Unless otherwise specified, the remaining parameters are set to $P=30$ dBm and  $L_{{\rm B},k}=L_{\rm E}=3$. The noise power spectral density is assumed to be $-174$ dBm/Hz. {All results are obtained by averaging over $100$ random channel realizations \cite{wang2023beamfocusing}.}
{\color{black}
To perform a comprehensive performance comparison, we consider the following benchmark schemes: 1) {\emph{Two-stage algorithm without AN:} The hybrid beamformers are designed using the algorithm in Section \ref{Sec:conventional}, without incorporating AN.} 2) \emph{ZF with AN:} The analog beamformer is obtained via the proposed algorithm, while the digital precoder is designed using ZF. 3) \emph{ZF without AN:}   This scheme is similar to the ZF with AN approach, but without including AN. 4) \emph{Fully analog:} The analog beamformer is applied jointly to both the legitimate users and AN, with only power allocation being optimized. 5) \emph{Far-field wideband:} The hybrid beamformers are designed using the algorithm from Section \ref{Sec:conventional}, under the assumption of a purely far-field wideband system. 6) {\emph{Near-field wideband:} The hybrid beamformers are designed using the algorithm from Section \ref{Sec:conventional}, under the assumption of a purely near-field wideband system.} 
Specifically, we also adopt the fully digital approximation-based hybrid beamforming method in \cite{wang2023beamfocusing}, named conv. hybrid [14], for performance evaluation.}
\begin{figure*}[t]
    \centering

    \begin{minipage}[t]{0.245\textwidth}
        \centering
        \includegraphics[width=\linewidth]{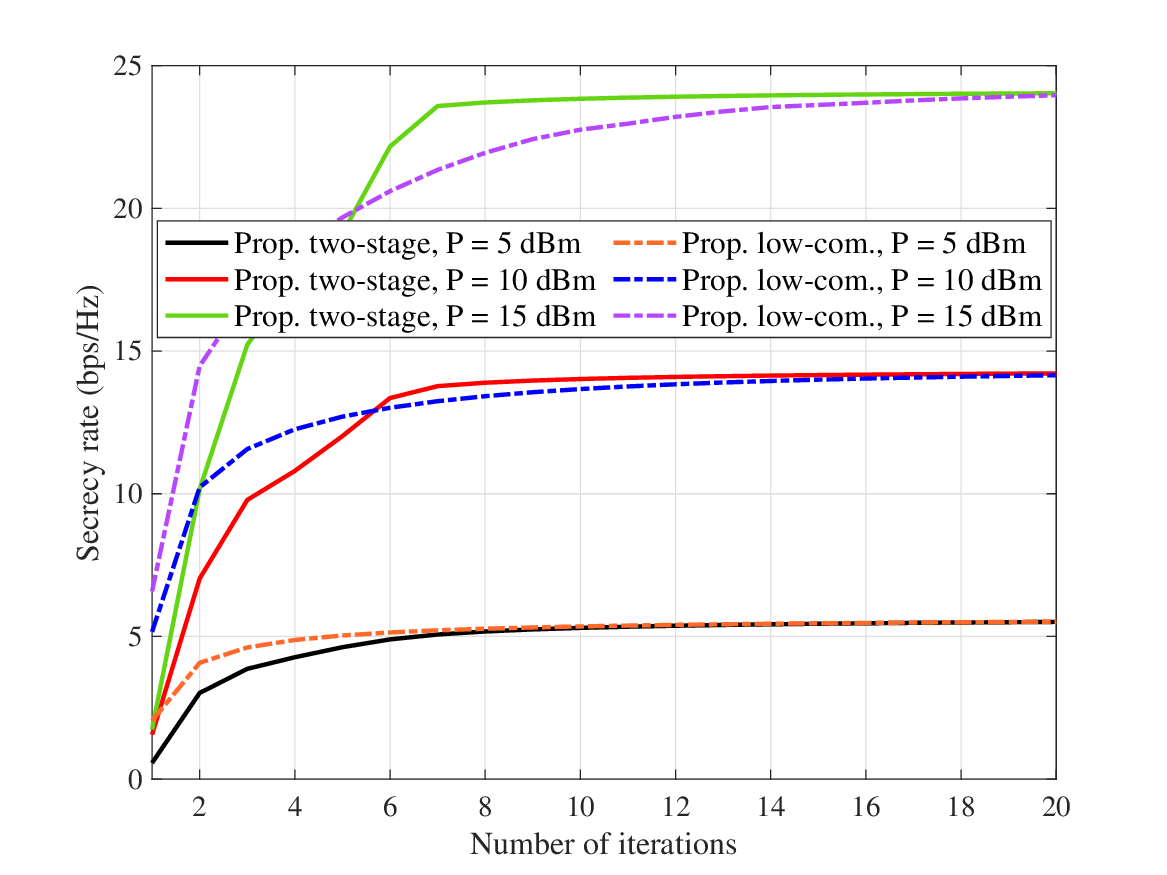}
        \vspace{-8pt}
        \captionof{figure}{\color{black}Secrecy rate versus iteration number.}
        \label{Fig:converagence}
    \end{minipage}
    \hfill
    \begin{minipage}[t]{0.245\textwidth}
        \centering
        \includegraphics[width=\linewidth]{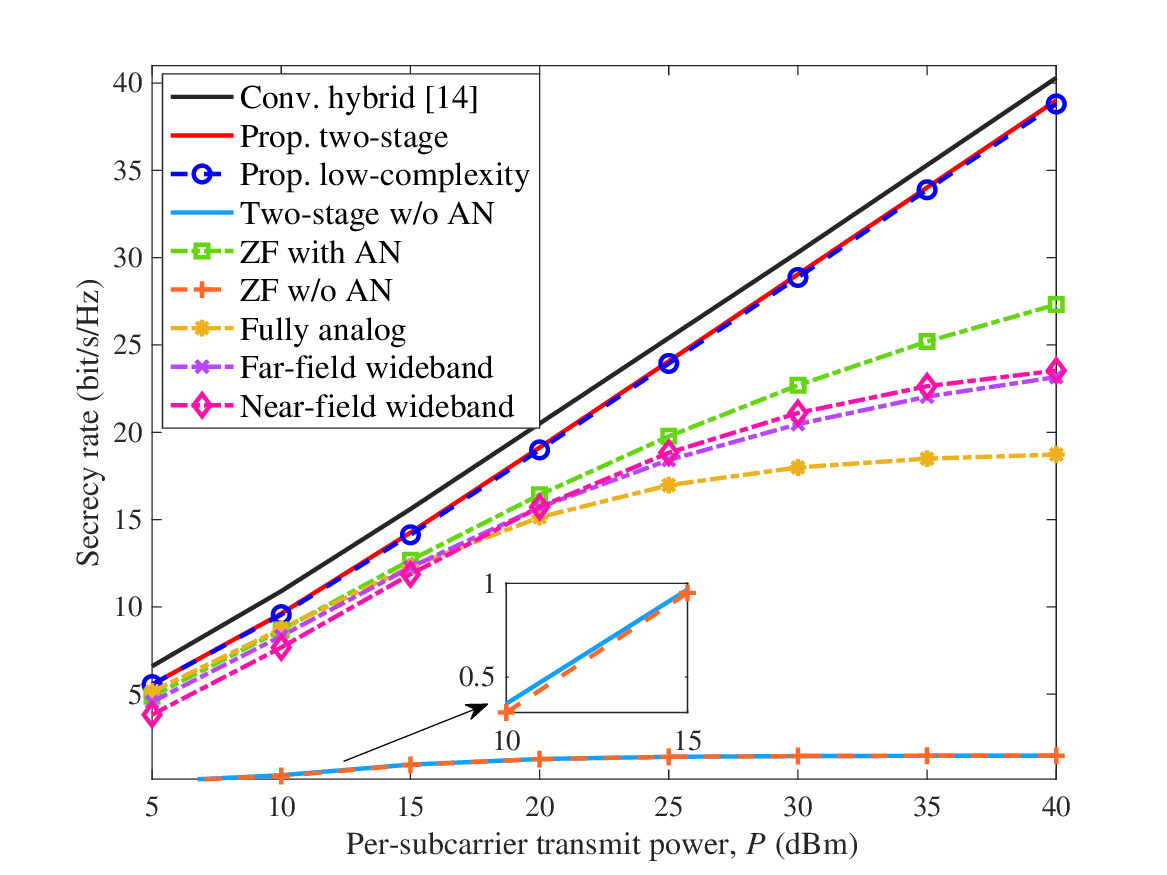}
        \vspace{-8pt}
        \captionof{figure}{\color{black}Secrecy rate versus per-subcarrier transmit power.}
        \label{Fig:secrecy_vs_power}
    \end{minipage}
    \hfill
    \begin{minipage}[t]{0.245\textwidth}
        \centering
        \includegraphics[width=\linewidth]{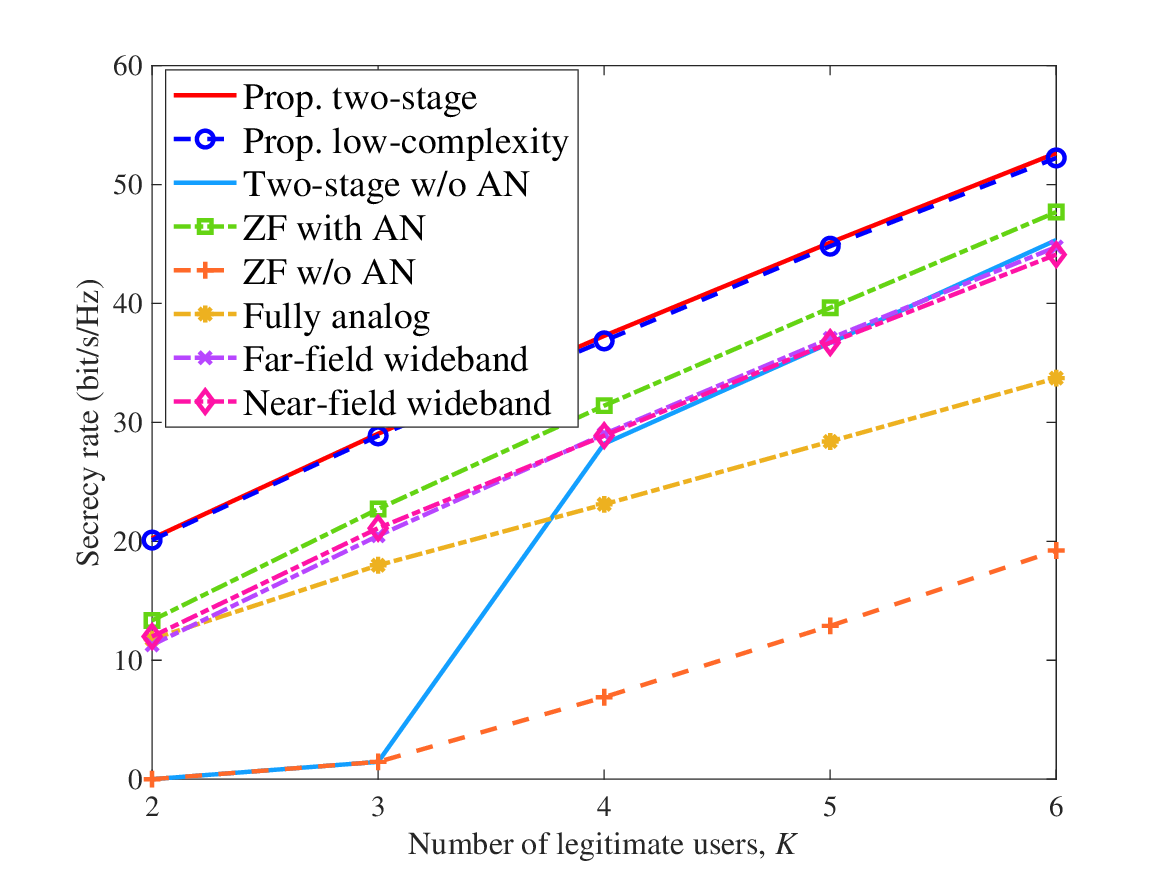}
        \vspace{-8pt}
        \captionof{figure}{\color{black}Secrecy rate versus number of legitimate users.}
        \label{Fig:secrecy_vs_numbob}
    \end{minipage}
    \hfill
    \begin{minipage}[t]{0.245\textwidth}
        \centering
        \includegraphics[width=\linewidth]{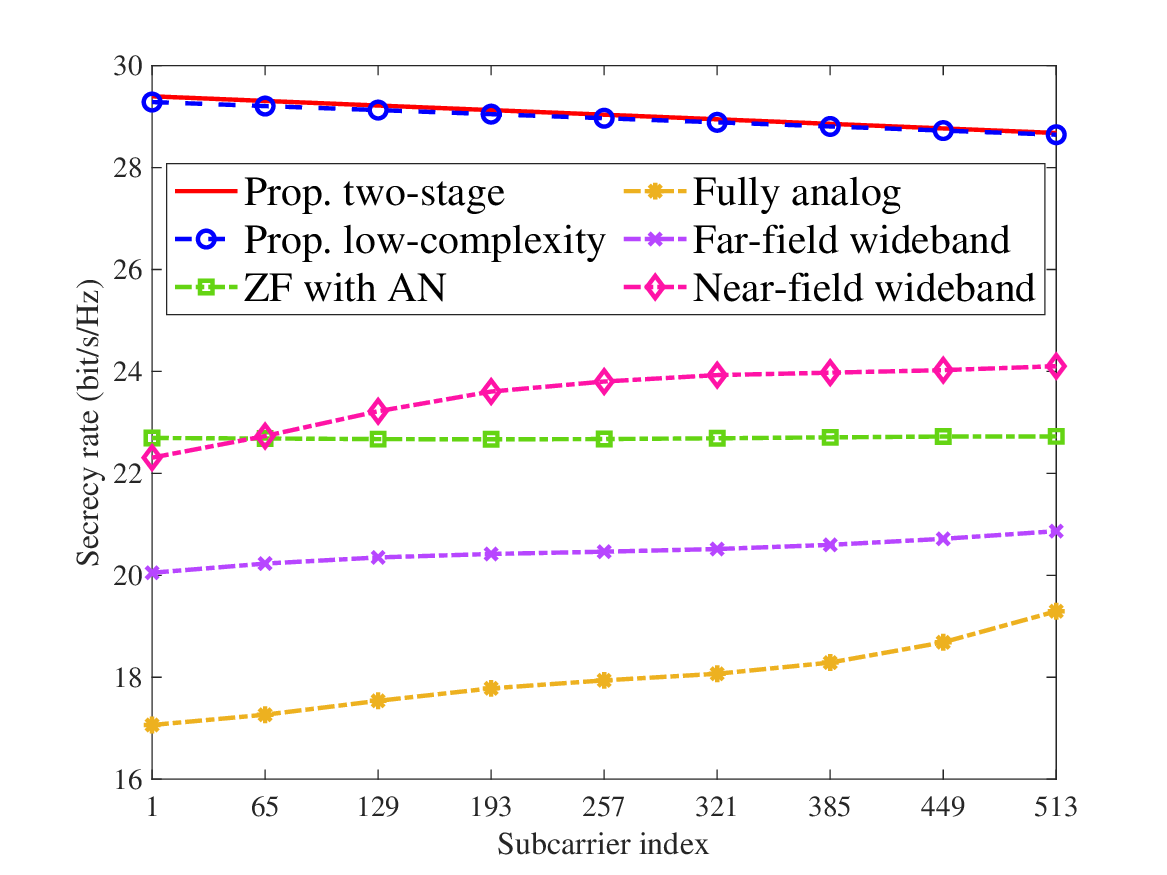}
        \vspace{-8pt}
        \captionof{figure}{\color{black}Secrecy rate versus subcarrier index.}
        \label{Fig:secrecy_vs_subcarrieridx}
    \end{minipage}
    \vspace{-15pt}
\end{figure*}
\vspace{-10pt}
\subsection{Algorithm Convergence}
In Fig. \ref{Fig:converagence}, we plot the secrecy rate versus the number of iterations to demonstrate the convergence behavior of the two proposed algorithms. It can be observed that both methods consistently converge within approximately $20$ iterations under varying per-subcarrier transmit power levels. Moreover, the low-complexity variant achieves performance very close to that of the two-stage algorithm.

% \begin{figure}[t]
% 	\centering
% \includegraphics[width=0.35\textwidth]{convergence.eps}
% 	\caption{\color{black}Secrecy rate versus iteration number.} 
%      \label{Fig:converagence}
%      \vspace{-15pt}
% \end{figure}
% \begin{figure}[t]
% 	\centering
% \includegraphics[width=0.35\textwidth]{secrecy_rate_vs_trans0620.eps}
% 	\caption{\color{black}Secrecy rate versus per-subcarrier transmit power.} \label{Fig:secrecy_vs_power} \vspace{-18pt}
% \end{figure}

% \begin{figure}[t]
% 	\centering
% \includegraphics[width=0.35\textwidth]{secrecy_rate_vs_numbob1.eps}
% 	\caption{\color{black}Secrecy rate versus number of legitimate users.} \label{Fig:secrecy_vs_numbob} \vspace{-15pt}
% \end{figure}
% \begin{figure}[t]
% 	\centering
% \includegraphics[width=0.35\textwidth]{secrecy_rate_vs_subcarrier3.eps}
% 	\caption{\color{black}Secrecy rate versus subcarrier index.} \label{Fig:secrecy_vs_subcarrieridx} \vspace{-18pt}
% \end{figure}

   \vspace{-10pt}
\subsection{Effect of Per-subcarrier Transmit Power}
In Fig. \ref{Fig:secrecy_vs_power}, we show the effect of per-subcarrier transmit power on the secrecy rate. It is evident that the secrecy rates achieved by all schemes increase monotonically with the transmit power. Notably, the proposed low-complexity algorithm achieves performance comparable to that of the two-stage algorithm. This highlights the importance of aligning the AN beamformer with the eavesdropper's channel, which is consistent with Remark \ref{Re:ANdesign} and confirms the analytical results presented in Section \ref{Sec:Special_case}. 
More importantly, both proposed algorithms substantially outperform the benchmark schemes, especially at higher transmit power levels, i.e., $P\ge20$ dBm.
In contrast, all other methods, except the two proposed algorithms, exhibit performance saturation as transmit power increases. The underlying reasons are as follows. For the two-stage scheme without AN, while inter-user interference is effectively suppressed, information leakage to the eavesdropper remains unmitigated. In the ZF with AN scheme, inter-user interference is suppressed at the expense of reduced array gain, which limits its performance. The ZF without AN scheme suffers from both array gain reduction and unmitigated information leakage, leading to significant secrecy degradation.
For the fully analog scheme, MRT-based beamforming limits the ability to exploit spatial-domain DoFs, rendering it ineffective in suppressing multi-user interference. The far-field wideband further reveals that neglecting the energy spread effect inherent in mixed-field wideband systems inevitably results in considerable performance degradation.
% Furthermore, the substantial performance gaps between schemes with and without AN, such as the proposed two-stage scheme versus its no-AN counterpart, or ZF with AN versus ZF without AN scheme, demonstrate the effectiveness of AN in improving mixed-field wideband PLS performance. 
% Similarly, the marked superiority of our approach over the ZF with AN scheme confirms the effectiveness of the two-stage hybrid beamforming framework. 

    \vspace{-10pt}
\subsection{Effect of Number of Legitimate Users}
In Fig. \ref{Fig:secrecy_vs_numbob}, we plot the secrecy performance versus the number of legitimate users. The secrecy rate for all schemes generally increases with the number of legitimate users. This is because each additional legitimate user contributes to a communication performance gain, despite the potential increase in information leakage. A notable observation is that the secrecy performance of the two-stage algorithm without AN starts at a relatively low level but improves sharply as the number of users increases up to $K=4$. This behavior is attributed to the fact that the design flexibility of the two-stage framework is proportional to the number of legitimate users. A larger number of users provides greater spatial DoFs, allowing more effective hybrid beamforming design and thereby enhancing wideband mixed-field PLS.
\vspace{-10pt}
\subsection{Effect of Subcarrier Index}
In Fig. \ref{Fig:secrecy_vs_subcarrieridx}, we illustrate the secrecy performance with respect to the subcarrier index. The achievable secrecy rates vary across subcarriers for all schemes. Specifically, for both proposed algorithms, the secrecy rate decreases monotonically with increasing subcarrier index. This trend is consistent with the findings in Cases 2 and 3 of Example \ref{Ex:AN} and Remark \ref{Re:two_conditions}, indicating that while information leakage is well suppressed, the reference path loss becomes the dominant factor. Since lower subcarrier indices correspond to lower frequencies and therefore smaller path losses, they yield higher secrecy. 
In contrast, the fully analog, far-field wideband, and near-field wideband schemes exhibit an increasing trend with the subcarrier index. This behavior arises because these schemes are unable to effectively suppress the information leakage, and secrecy performance is instead dominated by cross-correlation effects.
In such cases, higher subcarrier indices lead to weaker cross-correlations and thus lower information leakage, resulting in improved secrecy performance. This aligns with Case 1 in Example \ref{Ex:AN} and Remark \ref{Re:two_conditions}.

\begin{figure}[t]
	\centering
\includegraphics[width=0.4\textwidth]{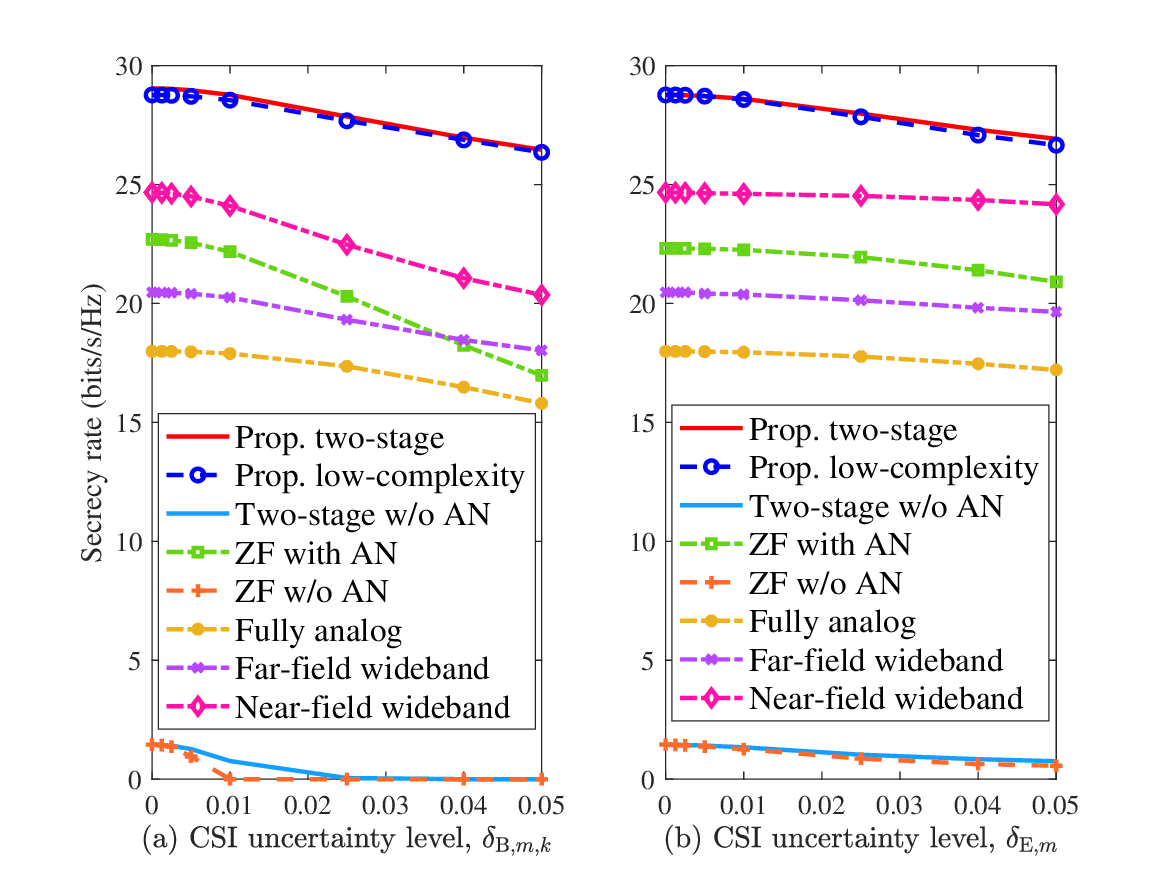}
	\caption{\color{black}Secrecy rate versus CSI error  with $P=30$ dBm, $K=3$, and $M=513$.} \label{Fig:secrecy_vs_csi}
        \vspace{-18pt}
\end{figure}
 \vspace{-12pt}
{\color{black}
\subsection{Effect of Imperfect CSI}\label{Sec:csierror}
In Figs. \ref{Fig:secrecy_vs_csi}(a) and \ref{Fig:secrecy_vs_csi}(b), we evaluate the secrecy performance of the proposed approaches and the benchmark schemes
under varying levels of channel uncertainty in the legitimate users’ channels and the eavesdropper’s channel, respectively. The uncertainties are characterized using the widely adopted statistical CSI
error model \cite{yunpusecure}. Specifically, the actual channels of the legitimate users and the eavesdropper on subcarrier $m$ are modeled as 
\begin{align}
 \mathbf{h}_{{\rm B},m,k} &=  \hat{\mathbf{{h}}}_{{\rm B},m,k}+\Delta\mathbf{h}_{{\rm B},m,k},\\
    \mathbf{h}_{{\rm E},m} &=  \hat{\mathbf{{h}}}_{{\rm E},m}+\Delta\mathbf{h}_{{\rm E},m},
\end{align}
where $\hat{\mathbf{{h}}}_{{\rm B},m,k}$ and $\hat{\mathbf{{h}}}_{{\rm E},m}$ denote the estimated channels of the legitimate user $k$ and the eavesdropper on subcarrier $m$, respectively, while $\Delta\mathbf{h}_{{\rm B},m,k}$ and $\Delta\mathbf{h}_{{\rm E},m}$ represent the corresponding estimation errors. Under the statistical CSI error model, these errors are assumed to follow the circularly symmetric complex
Gaussian (CSCG) distributions:
\begin{align}
    \Delta\mathbf{h}_{{\rm B},m,k}&\sim \mathcal{CN}(\mathbf{0},\boldsymbol{\Sigma}_{{\rm B},m,k} ),\\
     \Delta\mathbf{h}_{{\rm E},m}&\sim\mathcal{CN}(\mathbf{0},\boldsymbol{\Sigma}_{{\rm E},m} ),
\end{align}
with covariance matrices $\boldsymbol{\Sigma}_{{\rm B},m,k}=\delta^2_{{\rm B},m,k}\|\mathbf{h}_{{\rm B},m,k}\|_2^2\mathbf{I}_{N}$ and $\boldsymbol{\Sigma}_{{\rm E},m}=\delta^2_{{\rm E},m}\|\mathbf{h}_{{\rm E},m}\|_2^2\mathbf{I}_{N}$, where $\delta_{{\rm B},m,k}$ and $\delta_{{\rm E},m}$ quantify the CSI uncertainty levels.

It can be observed that the secrecy rates of all schemes decrease as the CSI uncertainty level increases. Nevertheless, the two proposed algorithms consistently outperform all benchmark schemes. Furthermore, we observe that the secrecy rate is more sensitive to CSI errors in the legitimate users’ channels than in the eavesdropper’s channel. This is because the legitimate-user CSI errors degrade both the beamforming gain at the intended receivers and the level of information leakage, whereas CSI errors in the eavesdropper’s channel primarily affect only the estimation of its received signal strength.
}

   \vspace{-10pt}
\section{Conclusion}\label{Sec:conc}
In this paper, we investigated PLS in mixed near-field and far-field wideband communication systems employing XL-arrays. We formulated an optimization problem to maximize the secrecy rate for far-field legitimate users against a near-field eavesdropper by jointly designing transmit beamforming with frequency-selective AN. Our analysis revealed that the secure transmission regions in mixed-field wideband systems differ across subcarriers, characterized by secrecy conditions derived from the Fresnel integrals. Furthermore, we theoretically proved that frequency-selective AN not only secures vulnerable subcarriers but also significantly boosts overall secrecy. For the general system setup, we developed an efficient two-stage hybrid beamforming algorithm and a low-complexity variant for joint beamforming and AN design. Numerical results confirmed the inherent variations in secrecy rates across subcarriers and demonstrated the superiority of the proposed algorithms over benchmark schemes. 
{\color{black}In future work, it is of significant interest to examine the joint impact of energy spread and beam squint in other mixed-field wideband systems. For example, their coupled effects can lead to more intricate inter-user interference patterns, thereby calling for more sophisticated interference-mitigation strategies.}
   \vspace{-8pt}
\begin{appendices}
  \section{}\label{App0}
   Let $A_1=\sqrt{\frac{d^2f_m\sin^2{\theta_{\rm E}}}{cr_{\rm E}}}$ and $A_2=\frac{f_md(\cos{\phi_{\rm B}}- \cos{\theta_{\rm E}})}{cA_1}$. Using the Riemann integral, we rewrite \eqref{Eq:NF_corre} as $\rho(f_m,\theta_{\rm E}, r_{\rm E},\phi_{\rm B})\approx\frac{1}{N}\left|\int^{\tilde{N}}_{-\tilde{N}}e^{j\pi(A_1n+A_2)^2}dn\right|$ \cite{zhang2023mixed,zhang2023swipt}. With the substitution $t = \sqrt{2}A_1n+\sqrt{2}A_2$, we obtain $\rho(f_m,\theta_{\rm E}, r_{\rm E},\phi_{\rm B})=\frac{1}{\sqrt{2}A_1N}\left|\int^{\sqrt{2}A_1\tilde{N}+\sqrt{2}A_2}_{-\sqrt{2}A_1\tilde{N}+\sqrt{2}A_2}e^{j\pi\frac{1}{2}t^2}dt\right|$.
    We define
  $\gamma_1 =\frac{f_m}{f_c}(\cos{\phi_{\rm B}}-\cos{\theta_{\rm E}})\sqrt{\frac{r_{\rm E}}{\frac{f_m}{f_c}d\sin^2{\theta_{\rm E}}}}$ and 
    $\gamma_2 = \frac{N}{2}\sqrt{\frac{\frac{f_m}{f_c}d\sin^2{\theta_{\rm E}}}{r_{\rm E}}}$.
    Then, we have 
 $
 \rho(f_m,\theta_{\rm E}, r_{\rm E},\phi_{\rm B}) \approx |G(\gamma_1,\gamma_2)| = \left|\frac{\hat{C}(\gamma_1,\gamma_2)+j\hat{S}(\gamma_1,\gamma_2)}{2\gamma_2}\right|$,
    where $\widehat{C}(\gamma_1,\gamma_2)=C(\gamma_1+\gamma_2)-C(\gamma_1-\gamma_2)$ and $\widehat{S}(\gamma_1,\gamma_2)=S(\gamma_1+\gamma_2)-S(\gamma_1-\gamma_2)$. This completes the proof.
     \vspace{-8pt}
    \section{}\label{App1}
    We begin by rewriting the secrecy rate in \eqref{Eq:SC_rate_expression1} as $R^{\rm Sec}_{{\rm B},m}\overset{(b)}{=}  
    \log_{2}\left( 1+\frac{A_2-A_3}{A_1+A_3} \right),$
where
\begin{align}
    A_1 &= \frac{P_{{\rm E},m}^2 N^2c^4 \rho^2(f_m)}{256 f_m^4 \pi^4 r_{\rm B}^2 r_{\rm E}^2} 
        + \frac{P_{{\rm E},m} Nc^2\rho^2(f_m) \sigma^2}{16 f_m^2 \pi^2 r_{\rm B}^2} \nn\\
        &
        \quad\quad\quad+ \frac{P_{{\rm E},m} Nc^2\sigma^2}{16 f_m^2 \pi^2 r_{\rm E}^2} 
        + \sigma^4, \nn\\
    A_2 &= (P - P_{{\rm E},m}) \left( \frac{P_{{\rm E},m} N^2c^4}{256 f_m^4 \pi^4 r_{\rm B}^2 r_{\rm E}^2} 
        + \frac{Nc^2 \sigma^2}{16 f_m^2 \pi^2 r_{\rm B}^2} \right), \nn\\
    A_3 &= (P - P_{{\rm E},m}) \rho^2(f_m) \left( \frac{P_{{\rm E},m} N^2c^4 \rho^2(f_m)}{256 f_m^4 \pi^4 r_{\rm B}^2 r_{\rm E}^2} 
        + \frac{Nc^2 \sigma^2}{16 f_m^2 \pi^2 r_{\rm E}^2} \right).\nn
\end{align}
$(b)$ holds by setting $\sigma^2_{{\rm B},m}=\sigma^2_{{\rm E},m} = \sigma^2$. 
To determine the condition under which $R^{\rm Sec}_{{\rm B},m}\ge0$ holds, we first find the roots of the following function:
\begin{align}
    \!\!\eta(P_{{\rm E},m}) &\triangleq A_2\!-\!A_3= -\alpha P_{{\rm E},m}^2 \!+\! (P \alpha - \beta) P_{{\rm E},m} \!+\! P \beta,
\end{align}
where 
   $ \alpha = \frac{N^2c^4 (1 - \rho^4(f_m))}{256 f_m^4 \pi^4 r_{\rm B}^2 r_{\rm E}^2}$ and $ 
    \beta = \frac{Nc^2 \sigma^2 (r_{\rm E}^2 -  r_{\rm B}^2\rho^2(f_m))}{16 f_m^2 \pi^2 r_{\rm B}^2 r_{\rm E}^2}.$
Clearly, $\eta(P_{{\rm E},m})$ is a quadratic function in $P_{{\rm E},m}$, and its discriminant is given by
\begin{align}
    \!\!\Delta\! \!=\!\! \left[\! \frac{Nc^2}{16 f_m^2 \pi^2 r_{\rm B}^2 r_{\rm E}^2}\!\! \left( \!\!\frac{P Nc^2 (1 \!-\! \rho^4(f_m))}{16 f_m^2 \pi^2}\!+\! \sigma^2 (r_{\rm E}^2 \!-\! \rho^2(f_m) r_{\rm B}^2) \!\right) \!\!\right]^2.\nn
\end{align}
The discriminant $\Delta$ is always  non-negative, indicating that the function $\eta(P_{{\rm E},m})$ admits two real roots. To derive the closed-form expressions of these roots, we consider the following two cases. \emph{1) Same Roots:} When $\Delta=0$, we have $r^2_{\rm E}- r^2_{\rm B} \rho^2(f_m) = -\frac{P Nc^2 (1 - \rho^4(f_m))}{16 f_m^2 \pi^2\sigma^2}. $
			Under this condition, $\eta(P_{{\rm E},m})$ has two identical roots, given by
			$P_{{\rm E},m}^{(1)}=P_{{\rm E},m}^{(2)}=P.$ \emph{2) Different Roots:} When $\Delta >0$, the function $\eta(P_{{\rm E},m})$ has two distinct real roots:	
					$P_{{\rm E},m}^{(1)}=\frac{16f_m^2\pi^2\sigma^2(r^2_{\rm B} \rho^2(f_m)-r^2_{\rm E})}{Nc^2(1-\rho^4(f_m))}$ and $
					P_{{\rm E},m}^{(2)}=P$.	
			Since $P_{{\rm E},m}^{(1)}$ is always smaller than $P_{{\rm E},m}^{(2)}$, we focus on the case $P_{\rm E}^{(1)}<P_{\rm E}^{(2)}$, which corresponds to $	r^2_{\rm E}- r^2_{\rm B} \rho^2(f_m)> -\frac{PNc^2(1-\rho^4(f_m))}{16f_m^2\pi^2\sigma^2}$.  
			Under this condition, $P_{\rm E}^{(1)}$ can be negative or positive. When $P_{\rm E}^{(1)}<0$, we have $r^2_{\rm E}- r^2_{\rm B} \rho^2(f_m)\ge 0$, implying that secure transmission is inherently guaranteed. When $P_{\rm E}^{(1)}>0$, i.e., $r^2_{\rm E}- r^2_{\rm B} \rho^2(f_m)< 0$, the value of $P_{\rm E}^{(1)}$ lies  between zero and $P_{\rm E}^{(2)}$, thus unveiling the minimum AN power required to achieve secure communication. Combining the above analyses yields the secure transmission condition in \eqref{Eq:sc_withAN}, completing the proof.	
        \vspace{-14pt}
        \section{}\label{app2}
         We define a new function, i.e., $\xi(P_{{\rm E},m}) = \frac{A_1+A_2}{A_2+A_3}$.
    To find its maximum, we compute its first-order derivative:
    \begin{equation}
        \zeta(P_{{\rm E},m}) \triangleq \frac{\partial \xi(P_{{\rm E},m})}{P_{{\rm E},m}} = \Upsilon_1 P_{{\rm E},m}^2+\Upsilon_2P_{{\rm E},m}+\Upsilon_3.
    \end{equation}
    The derivative $\zeta(P_{{\rm E},m})$ is a quadratic function with respect to $P_{{\rm E},m}$. Since the cross-correlation $\rho(f_m)$ satisfies $0<\rho(f_m)\le 1$, we have $\Upsilon_1<0$ and $\Upsilon_2<0$ under typical system conditions. Therefore, the behavior of $ \zeta(P_{{\rm E},m})$ is determined by the sign of $\Upsilon_3$. If $\Upsilon_3>0$, i.e., the condition in \eqref{Eq:sec_max} is satisfied, then it is shown that $\zeta(P_{\rm E})$ is initially positive for small $P_{{\rm E},m}$ and becomes negative as $P_{{\rm E},m}$ increases. This implies that the secrecy rate first increases and then decreases. Consequently, a unique maximum exists and is achieved at the root of $\zeta(P_{\rm E})$, given by $P_{\rm E}= -\frac{\Upsilon_2}{{2\Upsilon_1}}-\frac{\sqrt{{\Upsilon_2^2}-4{\Upsilon_1}\Upsilon_3}}{2\Upsilon_1}$
Conversely, if $\Upsilon_3\le 0$, then $\zeta(P_{\rm E})<0$ for all $P_{{\rm E},m}$, and the secrecy rate monotonically decreases with increasing AN power. In this case, the maximum is attained when no power is allocated to AN.
\end{appendices}
    \vspace{-10pt}
\bibliographystyle{IEEEtran}
\bibliography{Ref}
\end{document}

%% file: header.tex
\newtheorem{definition}{\emph{\underline{Definition}}}

\newtheorem{lemma}{\emph{\underline{Lemma}}}

\newtheorem{proposition}{\emph{\underline{Proposition}}}

\newtheorem{example}{\bf Example}
\newtheorem{remark}{\bf \emph{\underline{Remark}}}

\def\({\left(}
\def\){\right)}

\def\b0{{\mathbf{0}}}

\newcommand{\nn}{\nonumber}